\documentclass[11pt,a4paper]{article}
\usepackage[left=2.5cm,top=3cm,right=2cm,bottom=3cm]{geometry}
\usepackage{authblk}
\usepackage[T1]{fontenc}
\usepackage{amssymb,amsmath,geometry}
\usepackage{mathtools}
\usepackage{graphicx}
\usepackage{fancyhdr}
\usepackage{fancybox}
\usepackage{shadow}
\usepackage{empheq}
\usepackage{titlesec}
\usepackage{titletoc}
\usepackage{soul, color}
\usepackage{float}
\usepackage{shorttoc}
\usepackage{xcolor}
\usepackage{array}
\usepackage{tabularx}
\usepackage{booktabs}
\usepackage{rotating}
\usepackage{bigstrut}
\usepackage{aeguill}
\usepackage{changepage}

\usepackage{hyperref}
\def\sym#1{\ifmmode^{#1}\else\(^{#1}\)\fi}
\usepackage{caption}
\usepackage{blindtext}

\usepackage{caption}
\usepackage{subcaption}
\usepackage{mathtools}

\usepackage{graphicx}

\usepackage{enumerate}

\usepackage{hyperref}
\hypersetup{backref, colorlinks=true, citecolor=blue!50!black}
\usepackage{graphicx}

\usepackage{titlesec}
\usepackage{setspace} 
\usepackage{indentfirst}

\makeatletter
  \newcommand\smalls{\@setfontsize\smalls{10.3pt}{6}}
\makeatother

\makeatletter
  \newcommand\footnotesizes{\@setfontsize\footnotesizes{9.6pt}{6}}
\makeatother

\usepackage[authoryear]{natbib}
\newsavebox\tmpbox

\usepackage{caption}
\usepackage{subcaption}

\usepackage{lineno}

\begin{document}



\title{The centripetal pull of climate: Evidence from European Parliament elections (1989–2019)}



\author[1,4,*]{Marco Due\~nas}
\author[2]{Hector Galindo-Silva}
\author[3,4]{Antoine Mandel}
\affil[1]{\small Institute of Economics -- Sant'Anna School of Advanced Studies, Pisa, Italy}
\affil[2]{\small Department of Economics, Pontificia Universidad Javeriana, Bogot\'a, Colombia }
\affil[3]{\small Centre d'Economie de la Sorbonne --  Paris School of Economics -- CNRS-Universit\'{e} Paris 1 Panth\'{e}on-Sorbonne, Paris, France}
\affil[4]{\small CLIMAFIN -- Climate Finance Alpha, Paris, France}
\affil[*]{Corresponding author email: marco.duenas@climafin.com}


\date{}


\maketitle
\vspace{-2pt}

\begin{abstract} 
\noindent 
This paper examines the impact of temperature shocks on European Parliament elections. We combine high-resolution climate data with results from parliamentary elections between 1989 and 2019, aggregated at the NUTS-2 regional level. Exploiting exogenous variation in unusually warm and hot days during the months preceding elections, we identify the effect of short-run temperature shocks on voting behaviour. We find that temperature shocks reduce ideological polarisation and increase vote concentration, as voters consolidate around larger, more moderate parties. This aggregated pattern is explained by a gain in support of liberal and, to a lesser extent, social democratic parties, while right-wing parties lose vote share. Consistent with a salience mechanism, complementary analysis of party manifestos shows greater emphasis on climate-related issues in warmer pre-electoral contexts. Overall, our findings indicate that climate shocks can shift party systems toward the centre and weaken political extremes.
\end{abstract}



\clearpage
\newpage

\section{Introduction}

Climate change and climate policy have become a core issue in contemporary politics \citep{Dechezlepretre2025, fabel2025relationship}. Rising global temperatures and the increasing frequency of extreme weather events constitute external shocks that influence how societies organise politically \citep{CarletonHsiang2026, fisher2022politicisation, Dechezlepretre2025}. As climate change emerges as a key issue of political conflict, it influences both electoral behaviour and party strategies, thereby modifying the structure of democratic competition. Understanding how climate pressures affect political dynamics is key to designing effective climate policies and fostering broad-based social commitments to environmental action \citep{DoyleMcEachernMacGregor2015, Kotchen2024}.

The European context is a well-suited setting for studying this matter. European Parliament elections are held regularly across all member states under a common institutional framework, enabling systematic cross-national comparisons while ensuring consistency in electoral outcomes. At the same time, Europe exhibits substantial heterogeneity in historical climate conditions, partisan landscapes, and public attitudes toward climate policy, which makes it possible to study how similar climatic shocks generate different political responses. Moreover, environmental issues have long been salient in European politics, but their electoral importance varies widely across countries and over time \citep{Hoffmann2022NatureCC, Schworer2024}. 

In this paper, we investigate how temperature shocks affect regional vote outcomes in the European Parliament between 1989 and 2019, using high-resolution gridded climate data matched with party-level electoral outcomes aggregated at the NUTS-2 regional level. Our empirical strategy exploits plausibly exogenous variation in regional daily temperatures to identify the impact of anomalously warm and hot days on electoral behaviour. To capture non-linearities, we bin daily average temperatures and use the count of days in each bin as regressors, in agreement with standard approaches in the climate–economy literature \citep{DeschenesGreenstone2011, DeryuginaHsiang2014, DellJonesOlken2014}. This empirical design provides systematic evidence on the political consequences of short-run temperature shocks.

We document three main findings. First, unusually warm conditions reduce ideological polarisation and increase the concentration of votes among larger parties, consistent with a centripetal shift toward more moderate political actors. Second, when disaggregating these aggregate patterns by party family, we find that vote shares increase for liberal parties and, to a lesser extent, for social democratic parties, while declining for right-wing parties. Third, on the supply side, analysis of party manifestos shows that liberal and social democratic parties expand their climate-related content in response to unusually warm periods, whereas other party families do not.

A growing literature examines how weather shocks influence political behaviour, often through short-run mechanisms such as turnout or retrospective evaluations of incumbents. For instance, \citet{gomez2007republicans} show that rainfall on U.S. election days reduces turnout, disproportionately harming Democrats, while \citet{bechtel2011lasting} find that flooding before German local elections increased incumbent re-election prospects. Other studies emphasise how extreme events near elections shift the salience of environmental issues or trigger punishment of governments perceived as mishandling crises \citep{apergis2021role}. These highlight immediate channels, whereas our focus is on medium-term effects: how climatic shocks in the year preceding elections reshape party competition.

A smaller but expanding strand analyses voting directly in response to climate shocks. Evidence ranges from ballot initiatives in North America \citep{KahnMatsusaka1997, ListSturm2006, stokes2016} to wildfire exposure in California \citep{HazlettMildenberger2020} and fracking-induced earthquakes in Oklahoma and Texas \citep{Boomhower2024}. In Europe, studies show that personal experience of climate change or natural disasters can raise support for Green parties \citep{Hoffmann2022NatureCC, HilbigRiaz2024}, though extreme events do not systematically increase parties’ environmental attention in their communications \citep{Wappenhansetalt2024}. Our paper complements this literature by focusing on European Parliament elections and examining how temperature shocks affect outcomes such as vote concentration and polarisation.

Another line of research links climate shocks to labour markets and partisan alignment. \citet{Bombardinietal2025NBERw34120} show that extreme temperatures and shifts in “green” versus “brown” employment---where green jobs are in industries that protect the environment or conserve resources, and brown jobs are in fossil-fuel industries---move U.S. vote shares toward Democrats, implying that climate change will tilt politics toward pro-climate legislation by 2050. In this paper, we provide a European counterpart: anomalously warm days reduce polarisation and boost support for liberal and, to a lesser extent, social democratic parties, while diminishing support for right-wing parties. These findings connect to a broader body of work showing that climate change has become a salient axis of political competition: in the U.S., partisan polarisation strongly shapes climate beliefs and policy preferences \citep{mccright2011politicization, bohr2017hot, smith2024polarisation}, and in Europe, parties adjust their climate positions in response to electoral incentives, populist framing, and competition from Greens \citep{carter2013greening, carter2021party, huber2021populism, jahn2021quick, jahn2022party}.

Finally, our study connects to the wider climate–economy literature, which documents the economic and social consequences of weather shocks, including effects on productivity, mortality, and income \citep{DeschenesGreenstone2011, DeryuginaHsiang2014, DellJonesOlken2014, Hsiang2016}. By linking these insights to electoral outcomes, we show that climate shocks have political consequences beyond economic channels or elite framing: they alter party strategies and reshape electoral competition by reducing polarisation and concentrating votes.

The remainder of the paper is organised as follows. Section \ref{section_empiricalstrategy} describes the data and sets out the empirical strategy. Section \ref{section_mainresults} presents the baseline results. Section \ref{section_mechanisms} explores mechanisms and considers alternative explanations and limitations. Section \ref{section_conclusion} concludes by summarising the main findings and discussing their implications for the study of climate change and political competition.

\section{Data and methods} 
\label{section_empiricalstrategy}

\subsection{Data}\label{datasection}

The first key dataset for this study is historical climate data for all regions in Europe. We draw on the Berkeley Earth Foundation dataset \citep[available at:][]{CopernicusC3S2021}, which provides high-resolution gridded daily observations of temperature and precipitation. This dataset combines land-based temperature analyses with an interpolated version of HadSST3 for sea surface temperatures. The data cover the period from 1880 to 2019 and are provided on a 1-degree latitude/longitude grid. The weather variables used in our analysis are average temperature, maximum temperature, and total precipitation. To construct regional-level measures, we aggregate the daily grid-level data at the NUTS-2 regional level by computing population-weighted averages. Population weights are based on grid-level population estimates, which are available at five-year intervals. Population data from 1980 to 1995 come from the Inter-Sectoral Impact Model Intercomparison Project (ISIMIP) \citep{ISIMIP.892838}, while data from 2000 onwards come from NASA's Gridded Population of the World (GPW) \citep{GPW_NASA}.

\begin{figure}[H]
\begin{center}
\caption{Temperature histograms}
\vspace{-0.3cm}
\begin{subfigure}{0.5\textwidth}
\caption{Temperature}
\includegraphics[width=8cm]{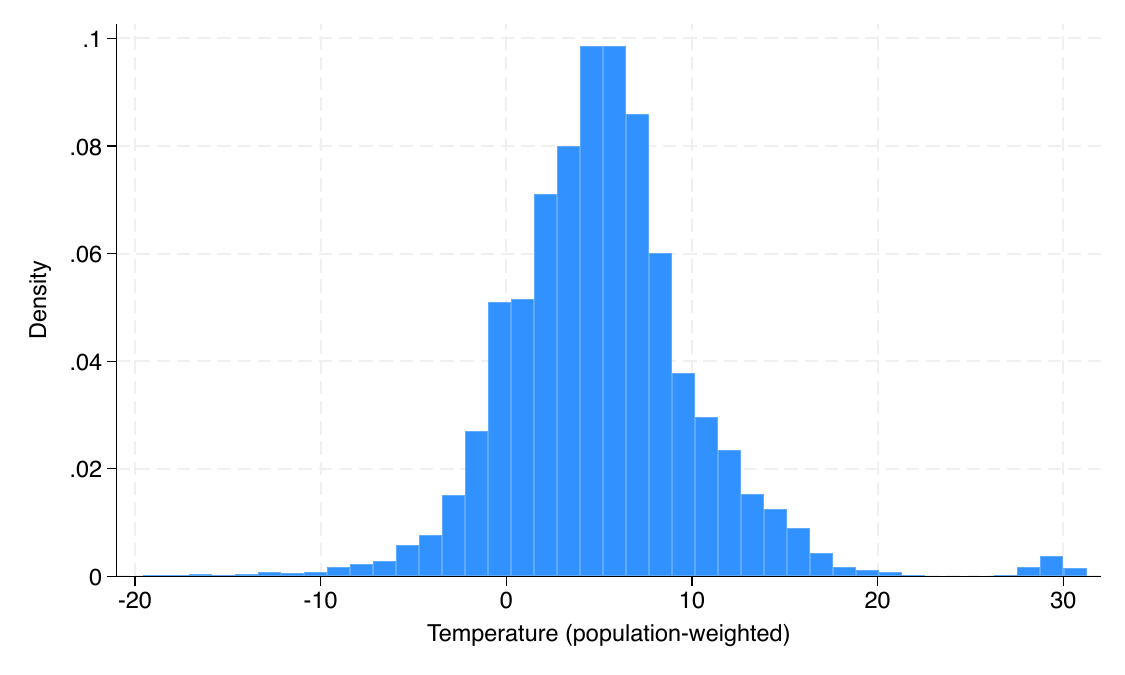}
\label{fig_temphistogram_a}
\end{subfigure}\hspace*{\fill}
\begin{subfigure}{0.5\textwidth}
\caption{Number of days}
\includegraphics[width=8cm]{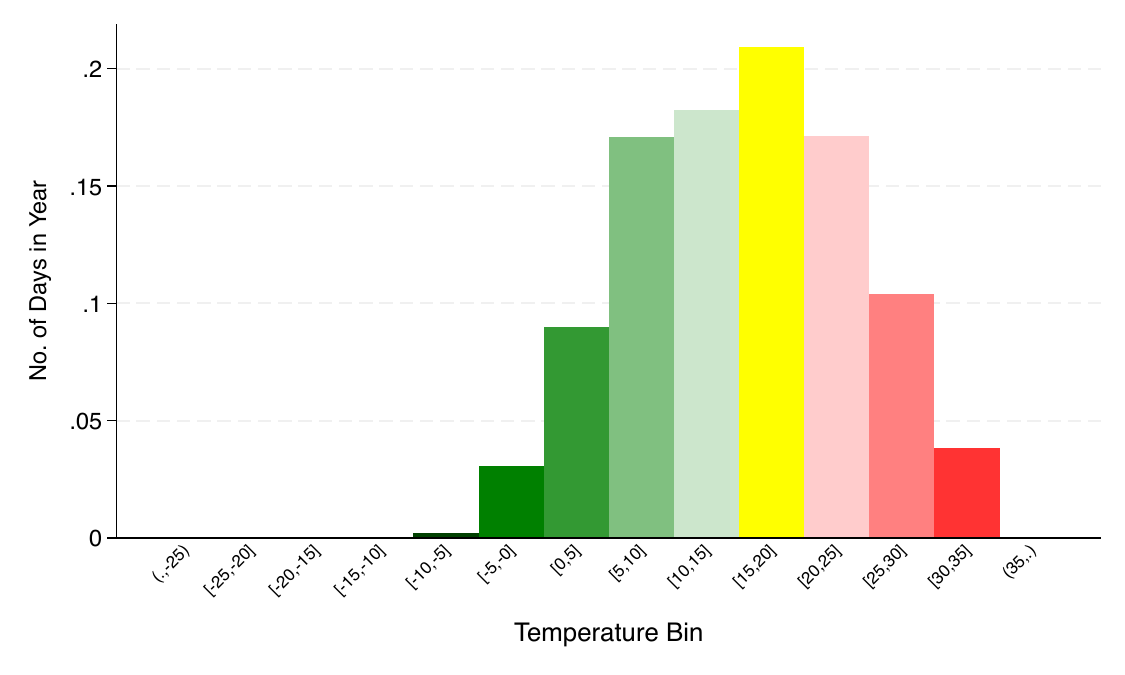}
\label{fig_temphistogram_b}
\end{subfigure}
\label{fig_temphistogram}
\begin{minipage}{0.9\textwidth} \scriptsize\textbf{Note}: Figure (a) presents the histogram of temperature distributions across 274 regions in Europe at the NUTS-2 level during the twelve months preceding the European Parliament elections. Figure (b) shows the distribution of the number of days in these same regions and periods across 12 temperature bins of 5$^{\circ}$C each, ranging from –35$^{\circ}$C to 35$^{\circ}$C. Two additional bins capture extreme values below –35$^{\circ}$C and above 35$^{\circ}$C.
\end{minipage}
\end{center}
\end{figure}

Figure \ref{fig_temphistogram_a} shows the distribution of temperatures across the full sample. Figure \ref{fig_temphistogram_b} presents a discrete version of the distribution of maximum daily temperatures across European regions. Daily temperatures are grouped into fourteen temperature categories or bins, each spanning $5^{\circ}$C intervals. The height of the bars represents the average number of days per year that the average person in each region experiences temperatures within each bin. Maps in Figure \ref{fig_map} in the Appendix display the spatial distribution of maximum daily temperatures at four points in time (January and July 1990 and 2018). These maps reveal substantial variation in temperature patterns across regions and over time.

The second dataset consists of electoral outcomes, drawn from the Parliaments and Governments Database (ParlGov) \citep{ParlGov}, which covers all EU democracies and most OECD democracies (37 countries) and provides information on political parties, elections, and cabinets from 1900 to 2023. Our analysis focuses on European Parliament elections held in 1989, 1994, 1999, 2004, 2009, 2014, and 2019.\footnote{In some countries, these elections took place in different years. Specifically, they were held in 1996 in Finland and Austria, in 2007 in Bulgaria and Romania, and in 2013 in Croatia. We include these cases in our main analyses. Key results remain robust when these are excluded.} We use party-level electoral outcomes aggregated to the NUTS-2 regional level to construct three measures: party vote concentration, party system ideological polarisation, and the regional vote share of different party families.

To measure party vote concentration, we use information on vote shares to compute the Herfindahl–Hirschman Index (HHI), defined as the sum of squared vote shares across all parties in a region and election:
\begin{equation}
\label{eq:hhi}
    HHI_i=\sum_j w_{ij}^2
\end{equation}
where $w_{ij}$ is the share of votes received by party $j$ in region $i$, computed for each electoral period.

We then incorporate ParlGov data on party positions, which are time-invariant, unweighted mean values of party ideology based on expert surveys, measured on a 0–10 left–right scale. We use this information to construct a measure of party system ideological polarisation, defined as the square root of the weighted sum of squared distances between each party’s ideological position and the system’s ideological centre of gravity (i.e., the weighted average party position), with party vote shares as weights \citep{Dalton2008}. Formally, polarisation is computed as:
\begin{equation}
\label{eq:polarisation}
Pol_i=\sqrt{\sum_{j} w_{ij}\Big(\frac{p_{j}-\overline{p}_i}{5}\Big)^{2}}  
\end{equation}
where $p_{j}$ denotes the left–right position of party $j$, $\overline{p}_i$ is the weighted average party position in region $i$, and $w_{ij}$ is the vote share of party $j$ in region $i$, again computed for each electoral period.

Finally, we use data on party families, which classify political parties based on their positions on both economic (state/market) and cultural (liberty/authority) left-right dimensions. This classification identifies eight party families: Communist/Socialist, Green/Ecologist, Social Democratic, Liberal, Christian Democratic, Agrarian, Conservative, and Right-Wing. Using this classification, we compute the vote share of each party family at the regional level.\footnote{Party families are typically classified by their ideological origins: social democrats emphasize welfare and equality, liberals favour limited government and market allocation, agrarians represent rural smallholders, Christian democrats stress a “Christian ethic,” and conservatives seek to preserve upper-class positions during periods of change (\citealp{Ziblatt2017, Delacerda2024}).} 

Table \ref{tab_descriptivestatistics} presents the descriptive statistics for the temperature variables, electoral variables, and other control variables, including regional population and the occurrence and intensity of extreme weather events. 

\subsection{Empirical strategy}

We analyse the impact of temperature on electoral outcomes by estimating a panel specification at the region-year level using a semi-non-parametric approach. Specifically, we consider the frequency with which daily temperature realisations fall into predefined bins (e.g., as in \citealt{DeschenesGreenstone2011} and \citealt{DeryuginaHsiang2014}) and estimate the following equation:
\begin{equation}
\label{baselinebins_eq}
E_{ict}=\theta_{i}+\theta_{ct}+\sum_{k}\beta_{k}TM^{k}_{ict-1} + \Gamma \textbf{X}_{ict-1}+\epsilon_{ict}
\end{equation}
where $E_{ict}$ represents the electoral outcome in region $i$ in country $c$ in election year $t$  (i.e., a year in which a European Parliament election was held: 1989, 1994, 1999, 2004, 2009, 2014, or 2019). The variables $TM^{k}_{ict-1}$ denote the number of days in region $i$ during the 12 months preceding the election year $t$ in which the maximum daily temperature fell within the temperature bin $k$. More precisely, the temperature distribution is divided into 12 bins of $5^{\circ}$C each, ranging from $-25^{\circ}$C to $35^{\circ}$C, with two additional bins capturing days with maximum daily temperatures below $-25^{\circ}$C and above $35^{\circ}$C, respectively (see Figure \ref{fig_temphistogram}). The reference temperature bin in all specifications is the $5^{\circ}$-$10^{\circ}$C bin, which has an equal number of bins on either side. Thus, the coefficient $\beta_{k}$ for temperature bin $k$ captures the effect of substituting a day in the reference bin with a day in bin $k$. 

The vector $\textbf{X}_{ict-1}$  includes indicator variables for rainfall during the 12 months preceding election year $t$, each capturing the presence of precipitation within one of six bins, ranging from less than 1 mm to more than 1031 mm. The first bin (excluded) serves as the reference category. The vector $\textbf{X}_{ict-1}$ also includes two time-varying variables measuring the presence (existence) and intensity (number of people affected) of natural disasters occurring during the 12 months preceding election year $t$.\footnote{We control for rainfall and extreme weather events to isolate the direct effect of temperature on electoral outcomes. As \cite{HoganSchlenker2024} note, however, this may not suffice, since other temperature-related factors---such as relative humidity, especially relevant for health outcomes---can also influence behaviour.}  In addition, $\textbf{X}_{ict-1}$ contains the natural logarithm of population and gross domestic product (GDP). Finally, $\epsilon_{ict}$ is the error term. To account for spatial correlation in this term, standard errors are clustered at the regional level in all specifications.

Equation \eqref{baselinebins_eq} also includes a full set of region fixed effects, $\theta_{i}$, which ensures that identification relies on within-region deviations from historical means. We interpret these yearly temperature shocks as random weather realisations within a given geographical area, and thus plausibly exogenous to electoral outcomes \citep{DellJonesOlken2014}.  Alternatively, these fixed effects absorb all unobserved, time-invariant regional determinants of electoral outcomes, meaning that factors such as latitude, altitude, land area, or historical adaptation strategies to climate variations do not confound the estimated effects of weather.

Additionally, the specification includes country-year fixed effects, $\theta_{ct}$, which account for time-varying factors common to all regions within a country, such as changes in national institutions, macroeconomic conditions (such as economic growth or recessions), electoral system characteristics, or large-scale migration flows.\footnote{\label{footnote_panelspec}As a complement to our baseline specification in Equation \eqref{baselinebins_eq}, we also estimate panel regressions of the form
$E_{ict}=\theta_{i}+\theta_{ct}+\sum_{j=0}^{L}\beta_{j}T_{ict-j}+\epsilon_{ict}$
where $T_{ict}$ denotes a vector of monthly average temperature and precipitation in region $i$, country $c$, and period $t-j$, with up to $L$ monthly lags. The identification assumptions mirror those of the baseline model. Because the effects of interest are plausibly non-linear, we present this specification only as a supplementary exercise rather than as our main empirical test.}

The validity of our main empirical strategy depends on the assumption that Equation \eqref{baselinebins_eq} provides unbiased estimates of the vector $\beta_{k}$. By conditioning on region and country-year fixed effects, $\beta_{k}$ is identified from deviations in regional weather relative to historical averages, while controlling for shocks common to all regions within a country. Given the unpredictability of short-term weather fluctuations, it is reasonable to assume that these variations are orthogonal to unobserved determinants of electoral outcomes.\footnote{The assumption that short-run weather variation is randomly assigned may fail if individuals access forecasts that enable ex ante adaptation. In that case, our estimates could conflate direct climate effects with forecast-driven adaptation. While some studies propose controlling for forecast information, this may exacerbate measurement error (see \citealp{CarletonDufloJackZappala2024Greenstone}), so we do not include such controls.}

\section{Effects of temperature shocks on electoral outcomes}
\label{section_mainresults}

We now present the estimated effects of temperature shocks on electoral outcomes, classifying the electoral outcomes into two categories: (i) measures of party vote concentration and party system ideological polarisation, and (ii) vote shares of each party family.

\subsection{Vote concentration and ideological polarisation}

\begin{figure}[H]
\begin{center}
\caption{Effect on party vote concentration and party system ideological polarisation}
\vspace{-0.3cm}
\begin{subfigure}{0.5\textwidth}
\caption{Party vote concentration}
\includegraphics[width=8cm]{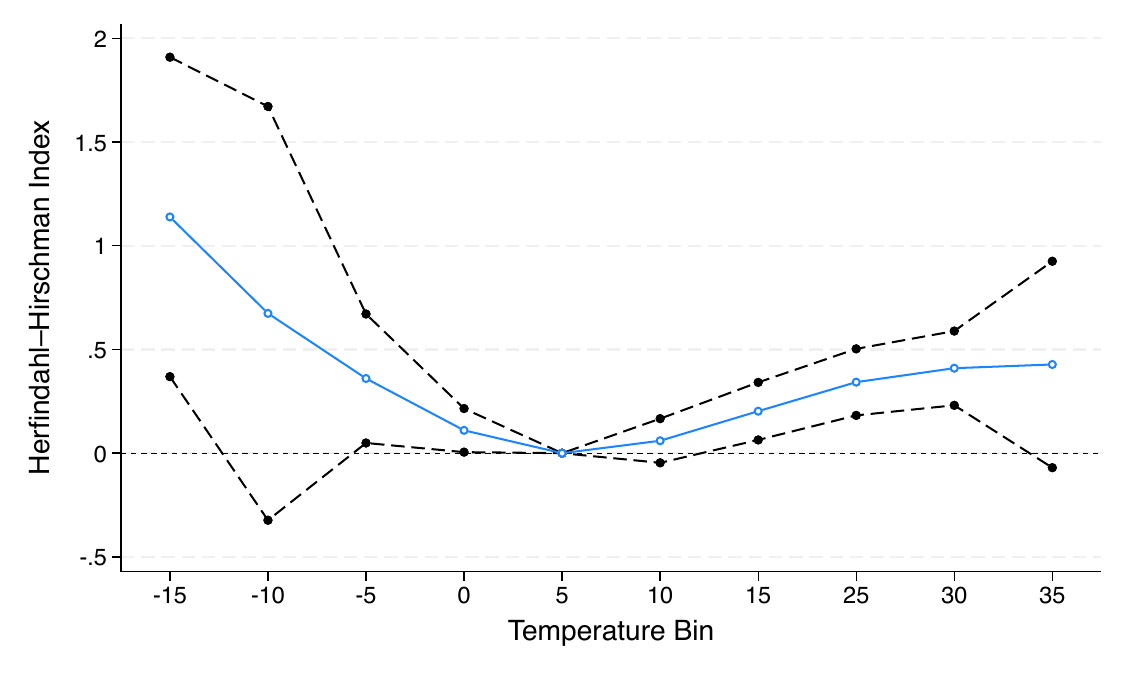}
\label{fig_HHI}
\end{subfigure}\hspace*{\fill}
\begin{subfigure}{0.5\textwidth}
\caption{Ideological polarisation}
\includegraphics[width=8cm]
{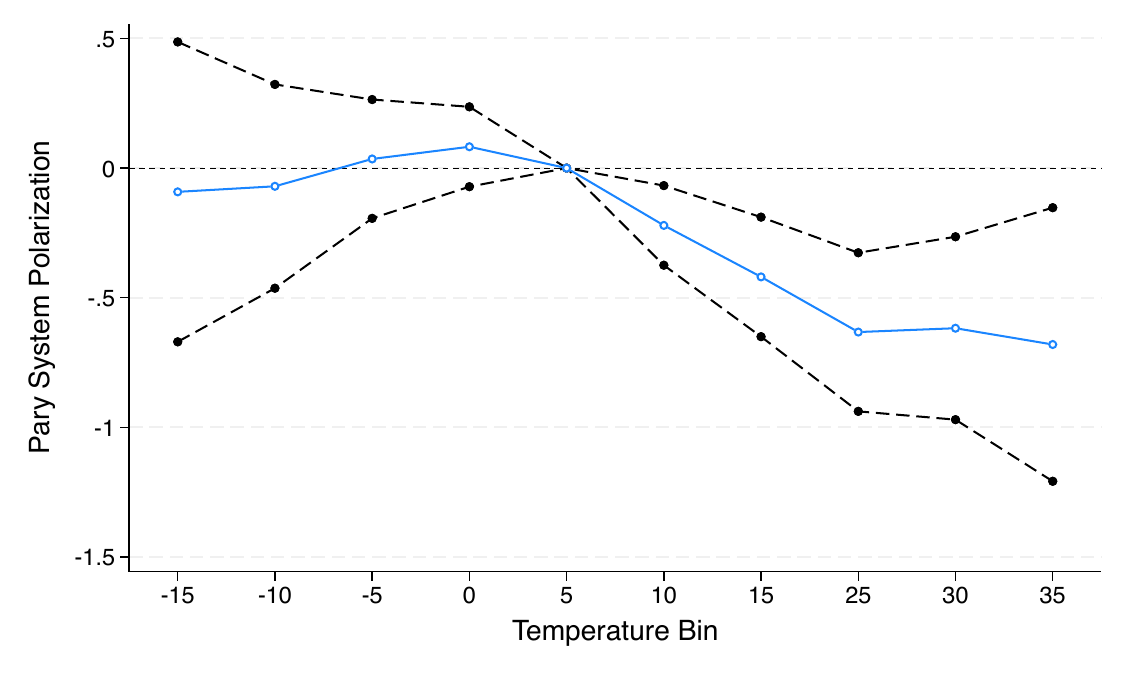}
\label{fig_polarisation}
\end{subfigure}
\label{fig_HHIpolarisation}
\begin{minipage}{0.9\textwidth} \scriptsize\textbf{Note}: The two figures report estimates from Equation \eqref{baselinebins_eq}, based on data from 274 European regions at the NUTS-2 level during the twelve months preceding the European Parliament elections. In Figure (a), the dependent variable is party vote concentration, measured by the Herfindahl–Hirschman Index (see Equation~\eqref{eq:hhi}). In Figure (b), the dependent variable is party system ideological polarisation (see Equation~\eqref{eq:polarisation}), computed for each electoral period. All models include region fixed effects, country–year fixed effects, and controls for precipitation, population, and both the number and intensity of extreme weather events. Shaded areas represent 95\% confidence intervals.
\end{minipage}
\end{center}
\end{figure}

Figure \ref{fig_HHIpolarisation} reports the estimates from Equation \eqref{baselinebins_eq}, using party vote concentration and party system ideological polarisation as dependent variables. For party vote concentration, Figure \ref{fig_HHI} shows increases when regions experience more days at both colder and hotter temperatures relative to the reference temperature bin $5^{\circ}$C–$10^{\circ}$C. For ideological polarisation, Figure \ref{fig_polarisation} shows declines at higher temperatures. These results are detailed in columns (2) and (4) of Table \ref{tab_turnHHIpol}, alongside results from specifications without controls (columns 1 and 3). 

For party vote concentration (columns 1 and 2, positive effects occur at lower temperatures (between $-15^{\circ}$C and $-10^{\circ}$C, and $-5^{\circ}$C and $5^{\circ}$C) and at higher temperatures (above $15^{\circ}$C). Within the $15^{\circ}$C–$35^{\circ}$C range, where effects are less noisy and more homogeneous, the estimated coefficients range from 0.2 to 0.4. This implies that replacing one day in the reference bin with a day in these warmer bins increases vote concentration by approximately 0.2–0.4 points. 

For ideological polarisation (columns 3 and 4), negative effects occur at higher temperatures (above $10^{\circ}$C), with coefficients ranging from $-0.22$ to $-0.68$. This indicates that replacing one day in the reference bin with a day in these warmer bins reduces ideological polarisation by about 0.22–0.68 points.\footnote{\label{footnote_extreme1}The results in Figure \ref{fig_HHIpolarisation} and Table \ref{tab_turnHHIpol} show that for extremely hot temperatures (above $5^{\circ}$C), the estimates—while similar in magnitude to those for less extreme temperatures—are noisier. A possible explanation for the larger standard errors in this bin is the very small number of regions that experienced at least one day in this temperature range, as illustrated in Figure \ref{fig_temphistogram_b}.}

These results reveal a consistent pattern: more days with higher temperatures are associated with reduced party vote fragmentation (via increased concentration) and less ideological opposition between major parties (via decreased polarisation). This pattern is consistent with a convergence of the vote toward the centre of the political spectrum, a mechanism we examine in more detail below.

\subsection{Vote share by party family}

\begin{figure}[H]
\begin{center}
\caption{Effect on vote share by party family}
\vspace{-0.3cm}
\begin{subfigure}{0.5\textwidth}
\caption{Communist/Socialist}
\includegraphics[width=7cm]{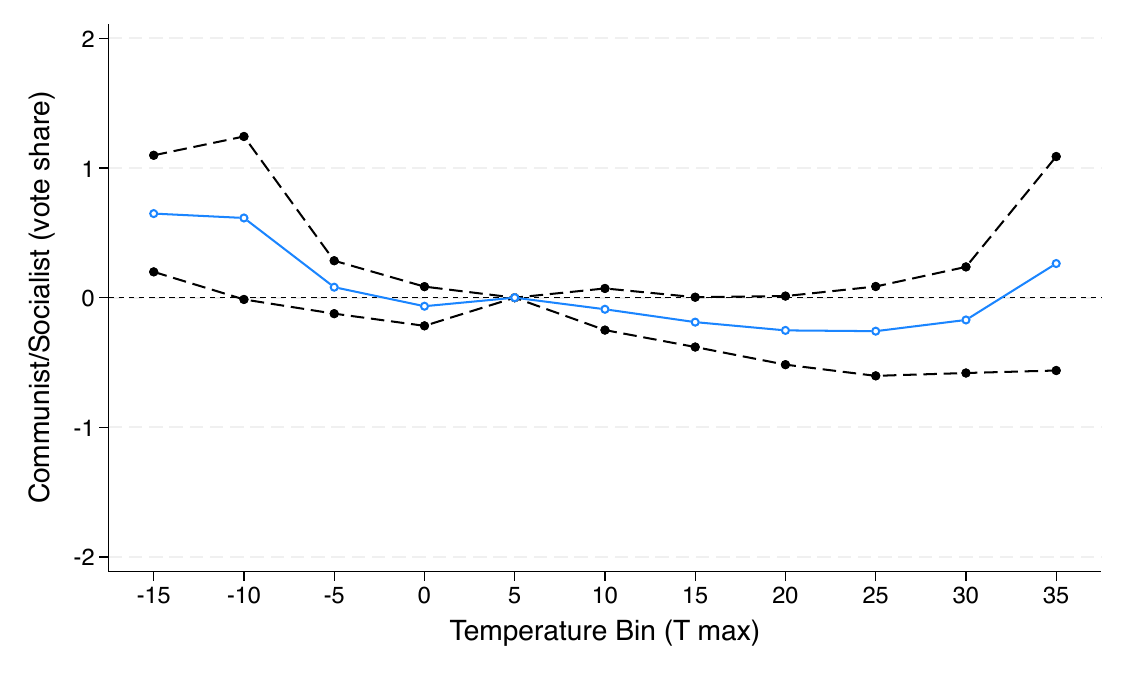}
\label{fig_families_g}
\end{subfigure}\hspace*{\fill}
\begin{subfigure}{0.5\textwidth}
\caption{Green/Ecologist}
\includegraphics[width=7cm]{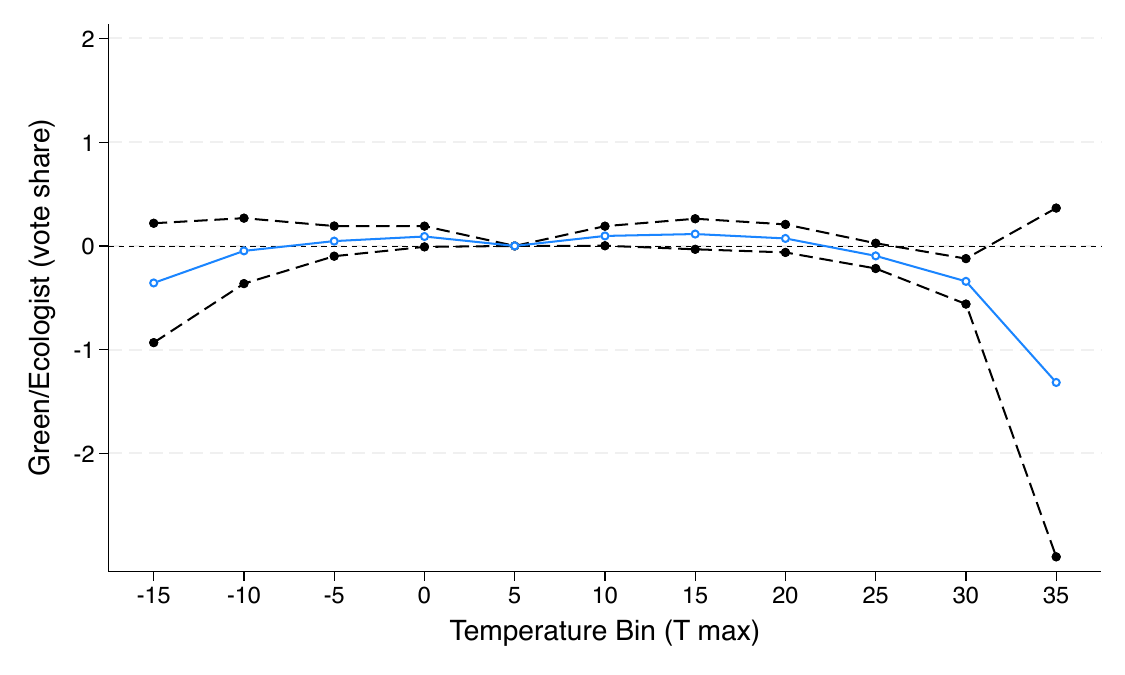}
\label{fig_families_b}
\end{subfigure}
\begin{subfigure}{0.5\textwidth}
\caption{Social Democratic}
\includegraphics[width=7cm]{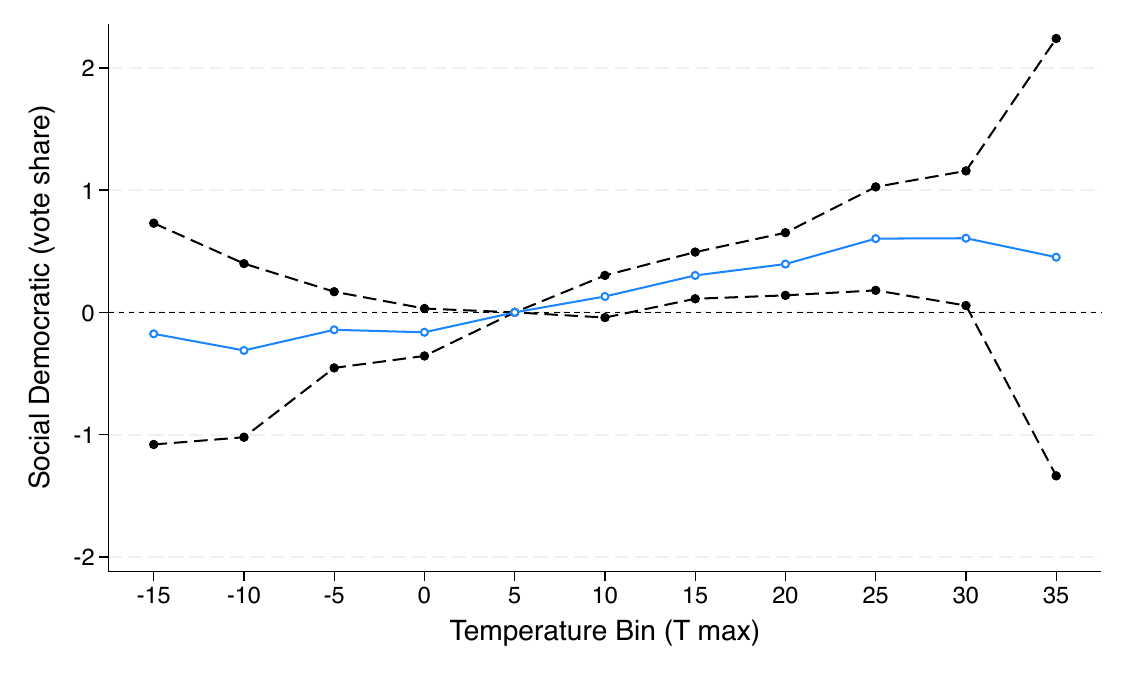}
\label{fig_families_e}
\end{subfigure}\hspace*{\fill}
\begin{subfigure}{0.5\textwidth}
\caption{Agrarian}
\includegraphics[width=7cm]{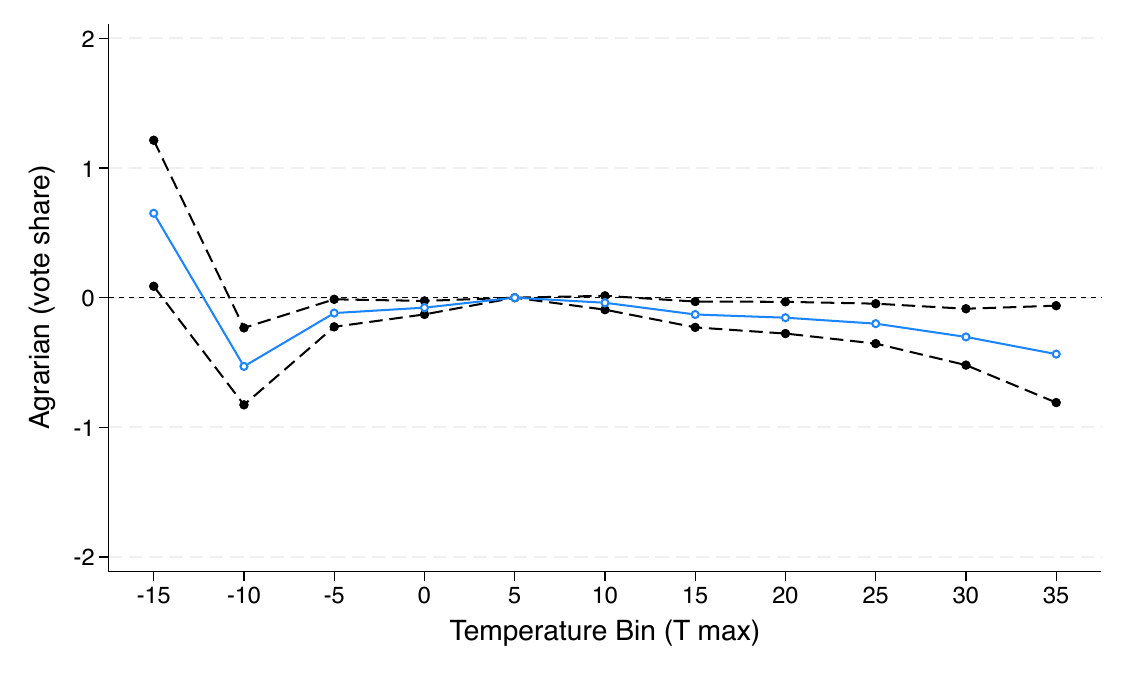}
\label{fig_families_h}
\end{subfigure}

\begin{subfigure}{0.5\textwidth}
\caption{Liberal}
\includegraphics[width=7cm]{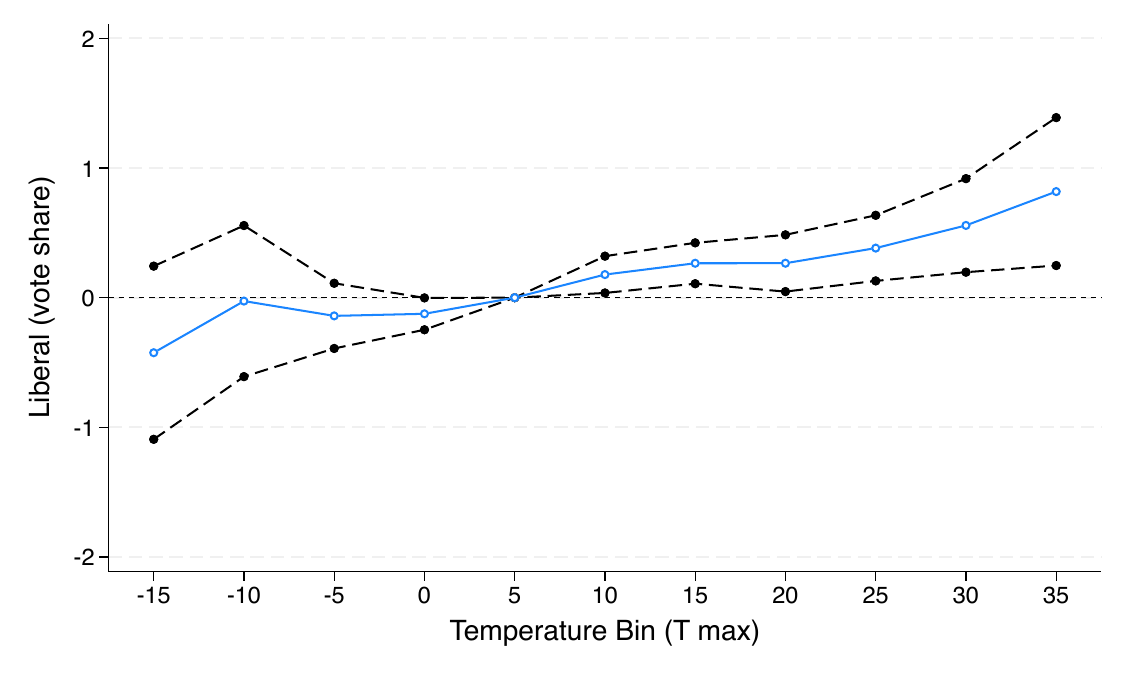}
\label{fig_families_c}
\end{subfigure}\hspace*{\fill}
\begin{subfigure}{0.5\textwidth}
\caption{Christian Democratic}
\includegraphics[width=7cm]{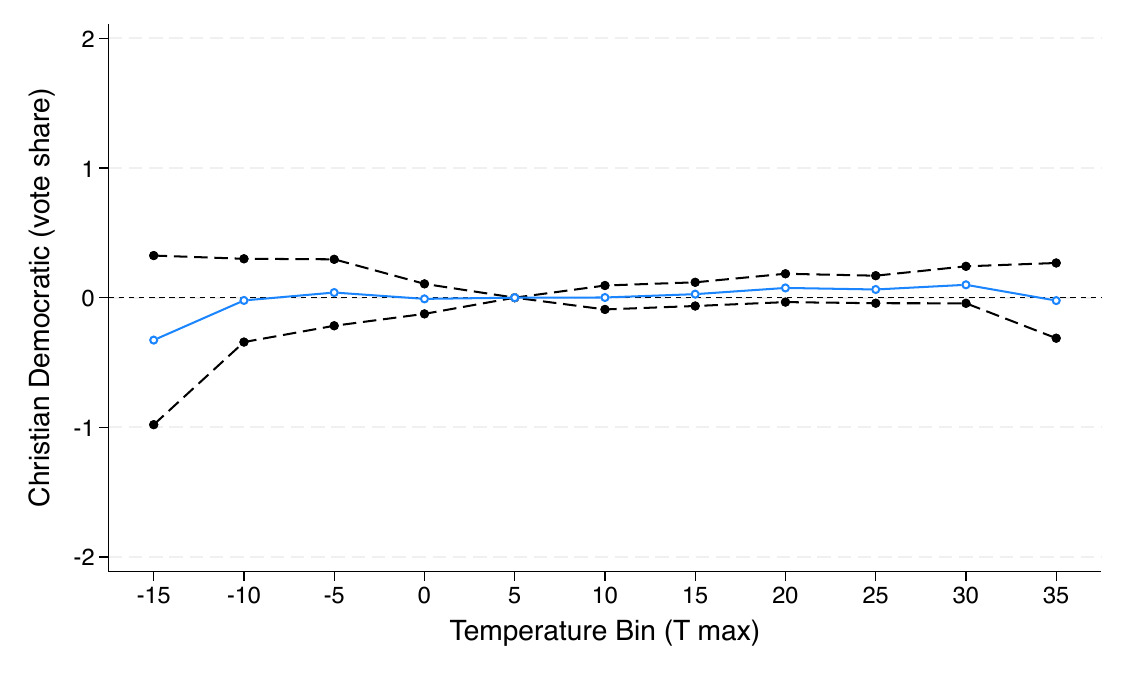}
\label{fig_families_a}
\end{subfigure}

\begin{subfigure}{0.5\textwidth}
\caption{Conservative}
\includegraphics[width=8cm,height=4cm]{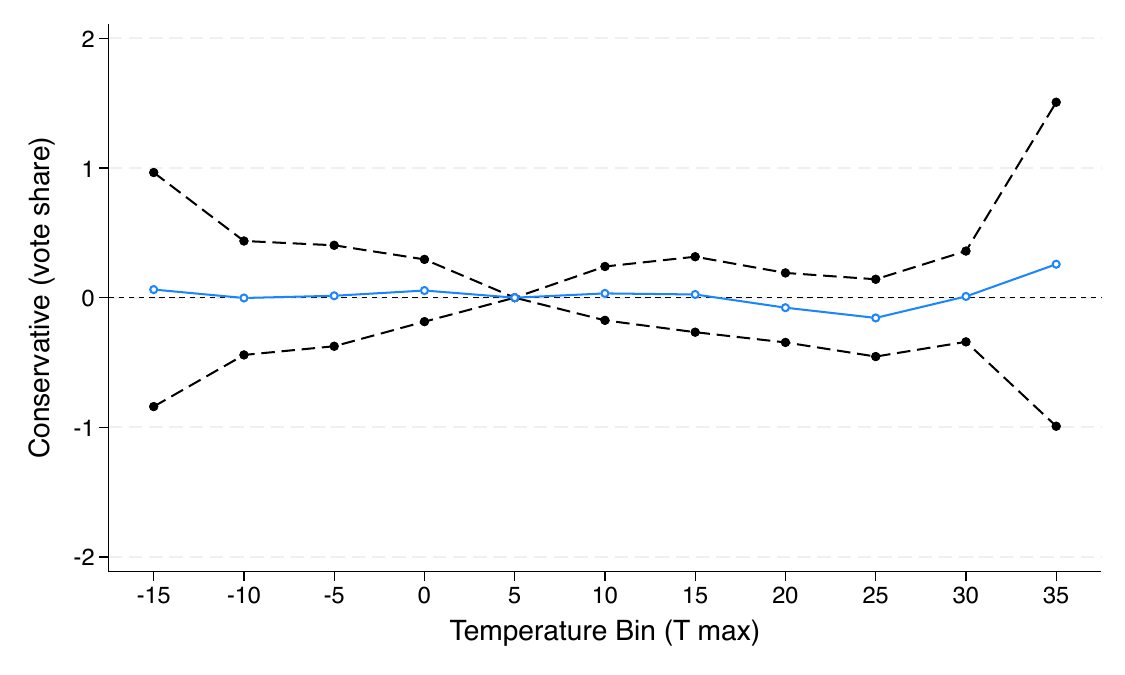}
\label{fig_families_f}
\end{subfigure}\hspace*{\fill}
\begin{subfigure}{0.5\textwidth}
\caption{Right-Wing}
\includegraphics[width=8cm,height=4cm]{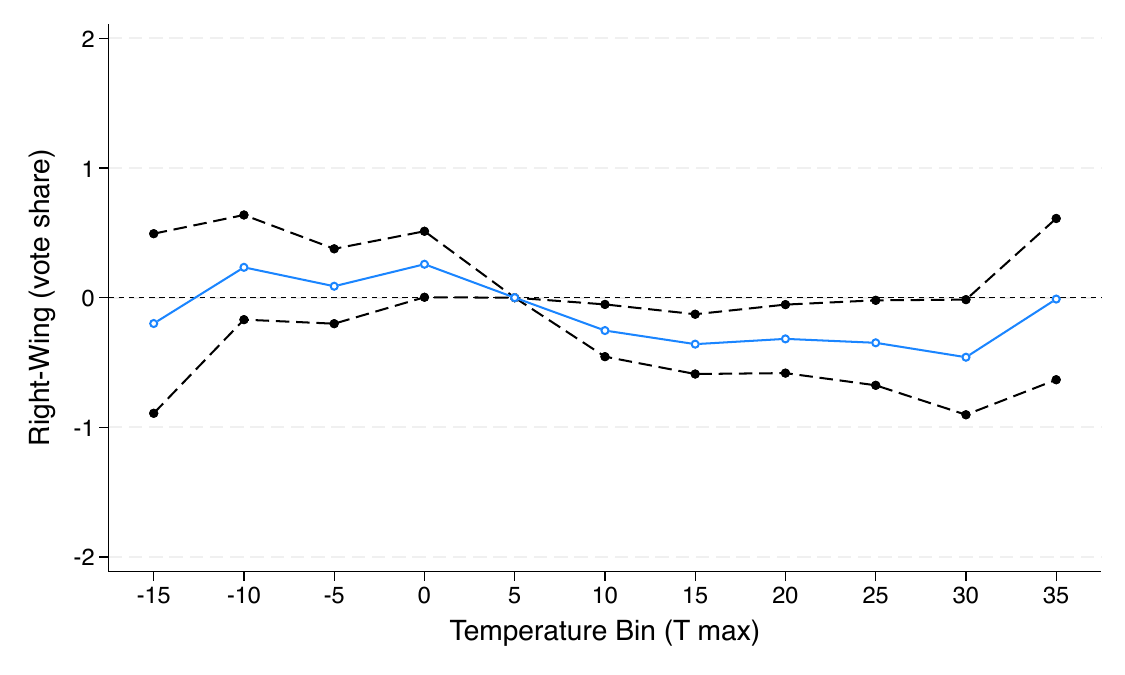}
\label{fig_families_d}
\end{subfigure}
\label{fig_families}
\begin{minipage}{0.9\textwidth} \scriptsize\textbf{Note}: All figures report the estimates from Equation \eqref{baselinebins_eq}, based on data from 274 regions in Europe at the NUTS-2 level during the twelve months preceding the European Parliament elections. The dependent variables in each figure are the vote shares obtained by the respective party families. All models include region fixed effects, country–year fixed effects, and controls for precipitation, population, and both the number and intensity of extreme weather events. Shaded areas indicate 95\% confidence intervals.
\end{minipage}
\end{center}
\end{figure}

Figure \ref{fig_families} and Table \ref{tab_idelologypartyfamily} report the estimates from Equation \eqref{baselinebins_eq}, using the vote share of each party family as the dependent variable. The results show statistically significant effects for certain party families. 

We find that more days with higher temperatures---above $10^{\circ}$C---are associated with increased vote shares for liberal and social democratic parties (Figures \ref{fig_families_c} and \ref{fig_families_e}; columns (3) and (5) of Table \ref{tab_idelologypartyfamily}). Coefficients range from 0.2 to 0.8 for both outcomes, implying that replacing a day in the reference bin with a warmer day increases their vote share by roughly 0.2–0.8 percentage points. For liberal parties, the effect peaks under extreme heat (0.8 points), whereas for social democratic parties, it is present only under moderately higher temperatures.

In addition, more days with higher temperatures are associated with declines in the vote share of right-wing and agrarian parties, with estimated decreases between 0.12 and 0.46 percentage points (Figures \ref{fig_families_d} and \ref{fig_families_h}; columns (4) and (8) of Table \ref{tab_idelologypartyfamily}). Additionally, extreme heat is linked to an increase in liberal party vote share (column 5) and a decrease in green party vote share (column 2).\footnote{\label{footnote_extreme2}Table \ref{tab_idelologypartyfamily} shows that extremely hot temperatures (above $35^{\circ}$C) are not associated with declines in the vote share of right-wing parties, in contrast to the negative effects observed for less extreme hot temperatures (column 8). A plausible interpretation is that when temperatures reach this range, climate change becomes highly salient, intensifying competition among relatively large parties. Liberal parties still gain vote share, but not at the expense of other large parties; instead, their gains come primarily at the expense of smaller ones. We develop this point further in the next section and present evidence consistent with this interpretation.} More days with lower---but not extreme---temperatures are associated with a decline in agrarian party vote share (column 4), while extremely low temperatures are associated with increases for communist/socialist parties and agrarian parties (columns 1 and 4). No statistically significant effects are found for other party families, such as Christian democrats and conservatives (Figures \ref{fig_families_a} and \ref{fig_families_f}; columns 6 and 7).\footnote{Appendix Table \ref{table_panelalloutcomes} presents the estimates obtained using the panel specification described in footnote \ref{footnote_panelspec}, restricted to the 12 months prior to the election. The results are broadly consistent with those reported in Figures \ref{fig_HHIpolarisation} and Table \ref{tab_idelologypartyfamily}, although in some cases they are estimated with less precision. In particular, the table shows a reduction in ideological polarisation (column 2), an increase in the vote share of liberal parties (column 7), and a decline in the vote share of right-wing parties (column 10). For vote concentration and the vote share of social democratic parties, the estimates have the same sign as in our baseline results but are statistically insignificant, which may suggest that non-linearities are more relevant for these outcomes.}

Overall, these findings are consistent with the earlier results on vote concentration and ideological polarisation: under warmer temperatures, relatively centrist and larger parties tend to gain votes at the expense of smaller and less centrist parties. In the next subsection, we show that these effects vary across regions when accounting for each region’s historical average temperatures. Nevertheless, the general patterns remain and constitute the main result of this paper. In the subsequent section, we propose an explanation for these findings.

\subsection{Heterogeneous effects by historic regional temperature}

The estimates above capture the average effects of unexpected temperature changes on electoral outcomes across European regions, irrespective of their historic average temperatures. Yet it is plausible that the effects vary systematically with historical temperature levels. For instance, the impact of an additional day between 10$^{\circ}$C and 15$^{\circ}$C (relative to 5$^{\circ}$C–10$^{\circ}$C) is likely quite different in regions with historic averages of --7$^{\circ}$C (e.g., Northern and Eastern Finland) compared to regions with averages of 16$^{\circ}$C (e.g., Southern Spain). To explore this heterogeneity, we re-estimate Equation \eqref{baselinebins_eq}, splitting the sample between regions with historic average temperatures (measured over 1980–1990) above and below 13.4$^{\circ}$C, the cross-regional median. 

\begin{figure}[H]
\begin{center}
\caption{Effect on party vote concentration and party system ideological polarisation by historic average temperature}
\vspace{-0.3cm}
\begin{subfigure}{0.5\textwidth}
\caption{Concentration \\ (above-median hist. temp)}
\includegraphics[width=8cm]{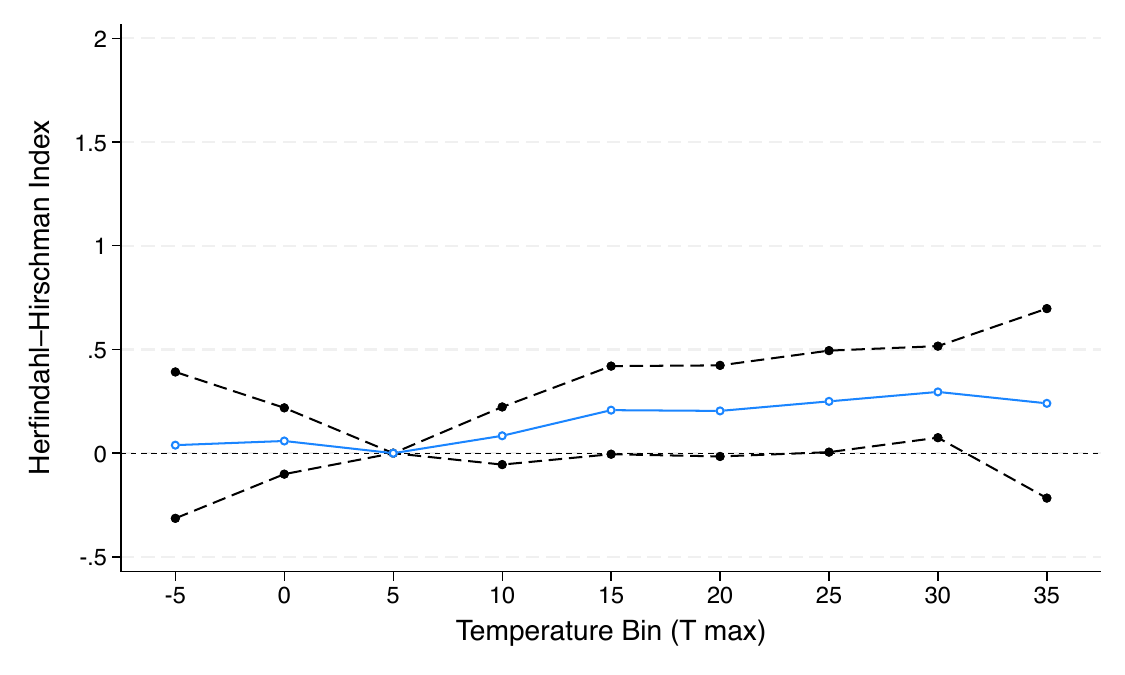}
\label{fig_histavgtemp_HHIPOL_a}
\end{subfigure}\hspace*{\fill}
\begin{subfigure}{0.5\textwidth}
\caption{Concentration \\ (below-median  hist. temp)}
\includegraphics[width=8cm]{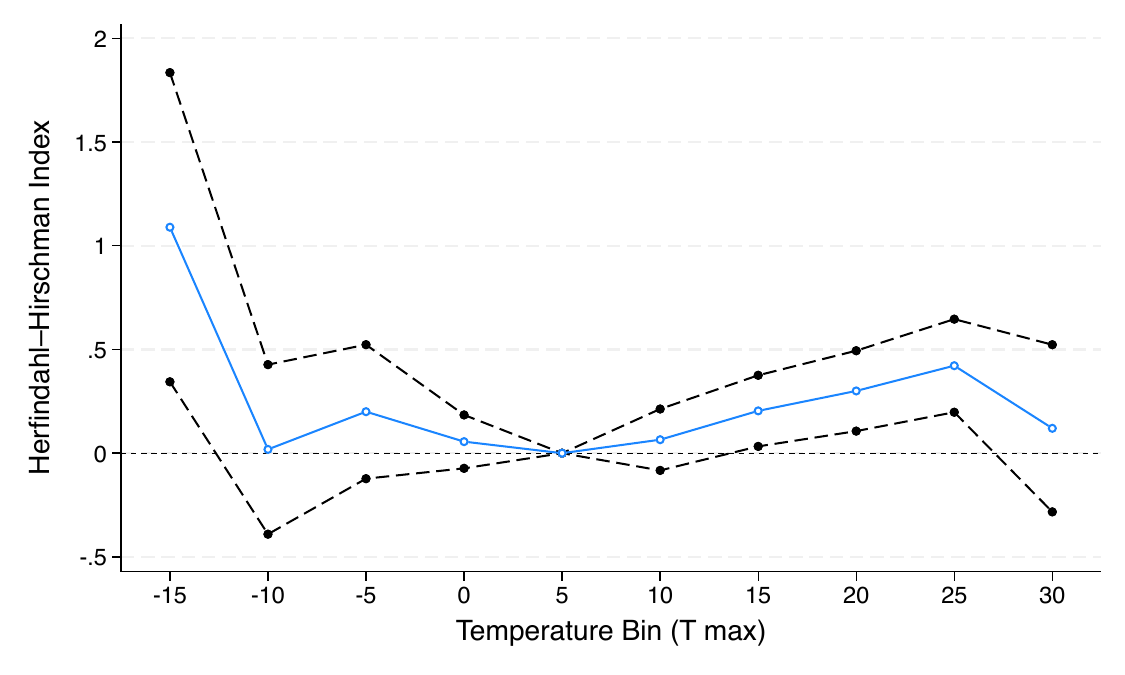}
\label{fig_histavgtemp_HHIPOL_b}
\end{subfigure}\\
\begin{subfigure}{0.5\textwidth}
\caption{Polarisation \\ (above-median hist. temp)}
\includegraphics[width=8cm]{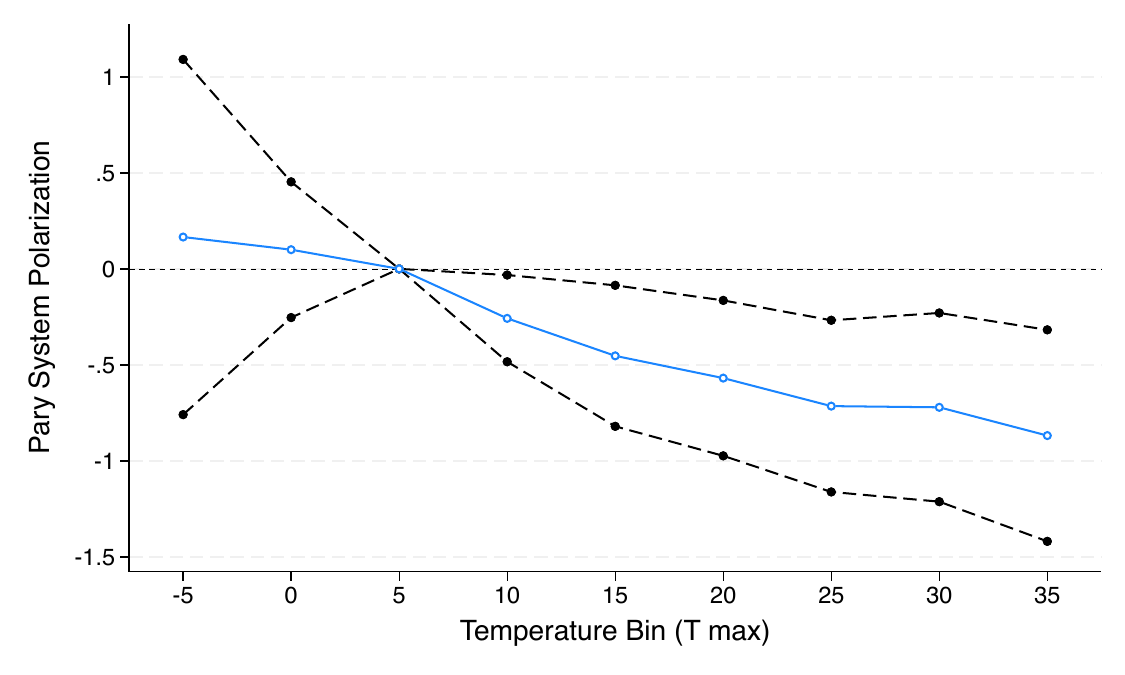}
\label{fig_histavgtemp_HHIPOL_c}
\end{subfigure}\hspace*{\fill}
\begin{subfigure}{0.5\textwidth}
\caption{Polarisation \\ (below-median  hist. temp)}
\includegraphics[width=8cm]{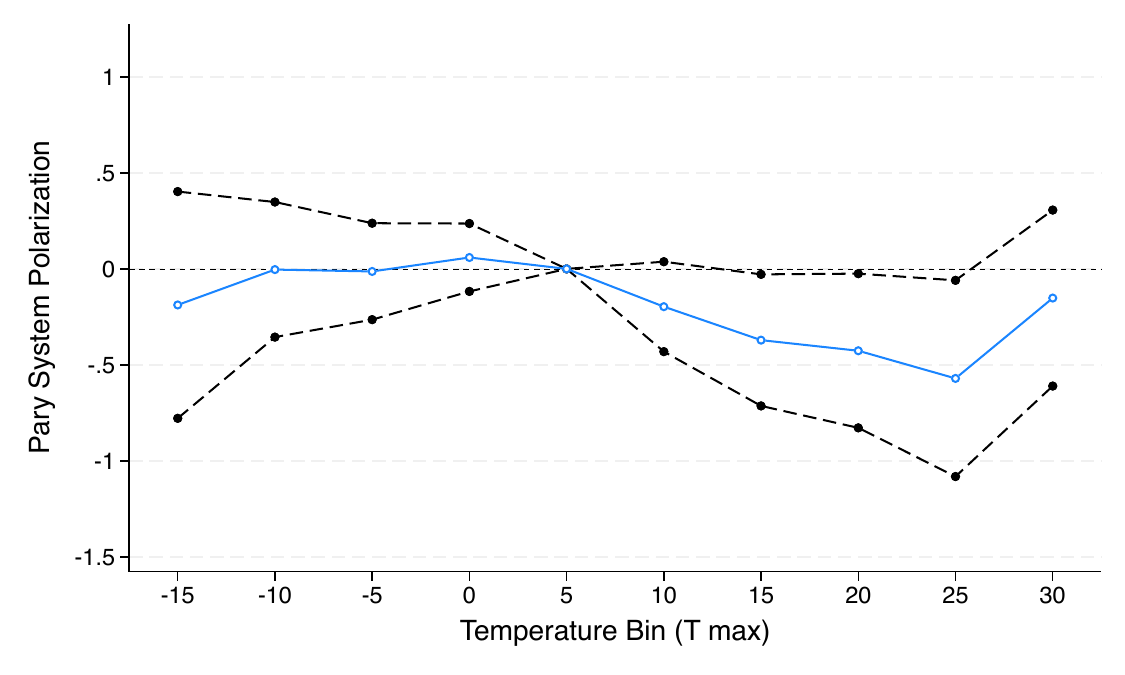}
\label{fig_histavgtemp_HHIPOL_d}
\end{subfigure}
\label{fig_histavgtemp_HHIPOL}
\begin{minipage}{0.9\textwidth} \scriptsize\textbf{Note}: All figures report estimates from Equation \eqref{baselinebins_eq}, using data from 274 European regions at the NUTS-2 level during the twelve months preceding the European Parliament elections. In Figures (a) and (b), the dependent variable is party vote concentration, measured by the Herfindahl–Hirschman Index (see Equation~\eqref{eq:hhi}). In Figures (c) and (d), the dependent variable is party system ideological polarisation (see Equation~\eqref{eq:polarisation}). Figures (a) and (c) are restricted to historically warm regions (above the median temperature), while Figures (b) and (d) are restricted to historically cold regions (below the median temperature). All models include region fixed effects, country–year fixed effects, and controls for precipitation, population, and both the number and intensity of extreme weather events. Shaded areas represent 95\% confidence intervals.
\end{minipage}
\end{center}
\end{figure}

Figure \ref{fig_histavgtemp_HHIPOL} reports the effects of temperature shocks on vote concentration and ideological polarisation. For vote concentration, significant effects---consistent with the baseline---are found in historically colder regions (Figure \ref{fig_histavgtemp_HHIPOL_b}), particularly for bins between 15$^{\circ}$C and 30$^{\circ}$C. In warmer regions, the effects are smaller and significant only between 25$^{\circ}$C and 35$^{\circ}$C (Figure \ref{fig_histavgtemp_HHIPOL_a}). For polarisation, the pattern is reversed: in warmer regions, effects are negative and significant at high temperatures (Figure \ref{fig_histavgtemp_HHIPOL_c}), whereas in colder regions they are weaker and significant only in a narrow range (15$^{\circ}$C–30$^{\circ}$C) (Figure \ref{fig_histavgtemp_HHIPOL_d}). Table \ref{Table_histavgtemp_HHIPOL} provides full estimates.

Party-level results offer further insight. In historically warmer regions (Table \ref{tab_idelologypartyfamily_high}), additional hot days between 15$^{\circ}$C and 35$^{\circ}$C increase liberal and social democratic vote shares (columns 5 and 3) and reduce right-wing support (column 8). At very high temperatures, green and agrarian parties also lose vote share. These shifts towards larger, centrist parties help explain the decline in polarisation and the modest rise in vote concentration.  

In colder regions (Table \ref{tab_idelologypartyfamily_low}), effects are noisier, but we find gains for Christian democratic and green parties and losses for conservative and agrarian parties, particularly at 15–20$^{\circ}$C (agrarian) and 25–30$^{\circ}$C (conservative). The decline in polarisation appears driven by increased support for centrist Christian democrats alongside a reduction in conservative votes. In these regions, higher vote concentration reflects the relative decline of smaller parties (agrarian, conservative, and to some extent right-wing) and gains for larger centrist parties. \footnote{In historically cooler regions, conservative parties average 14\% of the vote (versus 26\% in warmer regions), while Christian democrats average 20\% (versus 9\%). Liberal and right-wing parties have roughly similar support across both types of regions (10\% and 11–12\%).}

Taken together, the results indicate that unexpected hot days tend to reduce ideological polarisation and increase vote concentration, primarily by strengthening centrist and relatively large parties. In warmer regions, this occurs through gains for liberals and social democrats at the expense of right-wing and smaller parties, especially under extreme heat (above 30$^{\circ}$C). In colder regions, the pattern is driven by gains for Christian democrats and losses for conservatives at moderately high temperatures (20–30$^{\circ}$C). These differences are consistent with regional adaptation: the strongest and most systematic effects occur when shocks push temperatures beyond historically familiar ranges.


\section{Mechanisms}
\label{section_mechanisms}
\subsection{Climate as an electoral concern of political parties}

The results presented in the previous section suggest that certain political parties benefit electorally from the occurrence of specific types of temperature shocks. This pattern implies a reduction in ideological polarisation and an increase in vote concentration in the regions affected by such shocks. An important question arising from these findings concerns the mechanism through which some parties were able to increase their vote share as a consequence of unexpected temperature changes in the year preceding the election. In this section, we examine this question.

Our proposed explanation is as follows: temperature shocks increase the political salience of climate change among voters. Some parties decide to incorporate these concerns into their electoral platforms---and do so effectively. By taking such actions, they raise their chances of gaining electoral support.  

Testing this mechanism is challenging because we cannot directly observe how voters’ concerns evolve. Instead, we focus on the actions of political parties, for which we have more information. We assume that these actions reflect the interaction between voters’ demands and parties’ strategic decisions. 

The evidence we put forward in support of this explanation consists of several exercises. The first examines whether the effect of temperature shocks on the vote share of certain parties, as reported in the previous section, can be partly explained by the fact that parties gaining vote share in response to such shocks had performed poorly in the immediately preceding election. Our hypothesis is that this is indeed the case, and that the increase in vote share can be interpreted as the result of strategic adjustments made by these parties in response to prior electoral setbacks. 

Table \ref{tab_idelologypartyfamily_strenghtpast} shows the effect of temperature shocks on the vote share of liberal, social democratic, and right-wing parties---the same parties for which we previously found positive effects (liberals and social democrats) and negative effects (right-wing parties) in our baseline results in Table \ref{tab_idelologypartyfamily}. The results are now split between regions where liberals and social democrats had performed well in the previous election (i.e., above-median vote share; columns 1-3) and those where they had performed poorly (i.e., below-median vote share; columns 4-6). The table shows that nearly all effects are concentrated in regions where these parties had previously performed poorly: for liberals and social democrats, positive effects are found only in such regions (columns 4 and 5). Moreover, the negative effects for right-wing parties are also concentrated in these regions, consistent with the hypothesis that the results in Table \ref{tab_idelologypartyfamily} can be explained by the strategic reactions of liberals and social democrats to prior electoral underperformance. 

What did liberals and social democrats do to attract voters? Addressing this question is crucial because it is also possible that these parties did nothing, and instead, other parties may have taken missteps that explain the results. We approach this question by hypothesising that liberals and social democrats integrated references to climate change into their party platforms, which, in the context of hotter-than-expected temperature shocks, made them more relevant electoral options in subsequent contests. 

To test this hypothesis, we use data on references to climate change in party manifestos for a large number of European parties during the study period, collected by \cite{Schworer2024}. Specifically, we rely on \citeauthor{Schworer2024}’s measures for manifestos of 27 mainstream parties in 8 Western European countries, capturing the proportion of sentences dedicated to climate policy (climate protection salience) relative to other issues.\footnote{These data focus on large national parties with representation in national parliaments. This implies two limitations: (i) the sample does not include all party families; in practice, it covers Christian democrats, liberals, conservatives, and social democrats, so results are limited to these groups; (ii) the data are at the national level, preventing us from exploiting regional variation.} Using these data, we examine two questions: (i) whether systematic differences exist across party families in their references to climate change, and (ii) whether temperature shocks affect the frequency of such references. Our hypothesis is that liberal and social democratic parties include a higher proportion of climate change references in their manifestos, and that this proportion is influenced by the occurrence of temperature shocks.

\begin{figure}[H]
\begin{center}
\caption{References to climate change in party manifestos by party family of mainstream parties}
\vspace{-0.3cm}
\begin{subfigure}{0.5\textwidth}
\caption{Any reference}
\includegraphics[width=8cm]{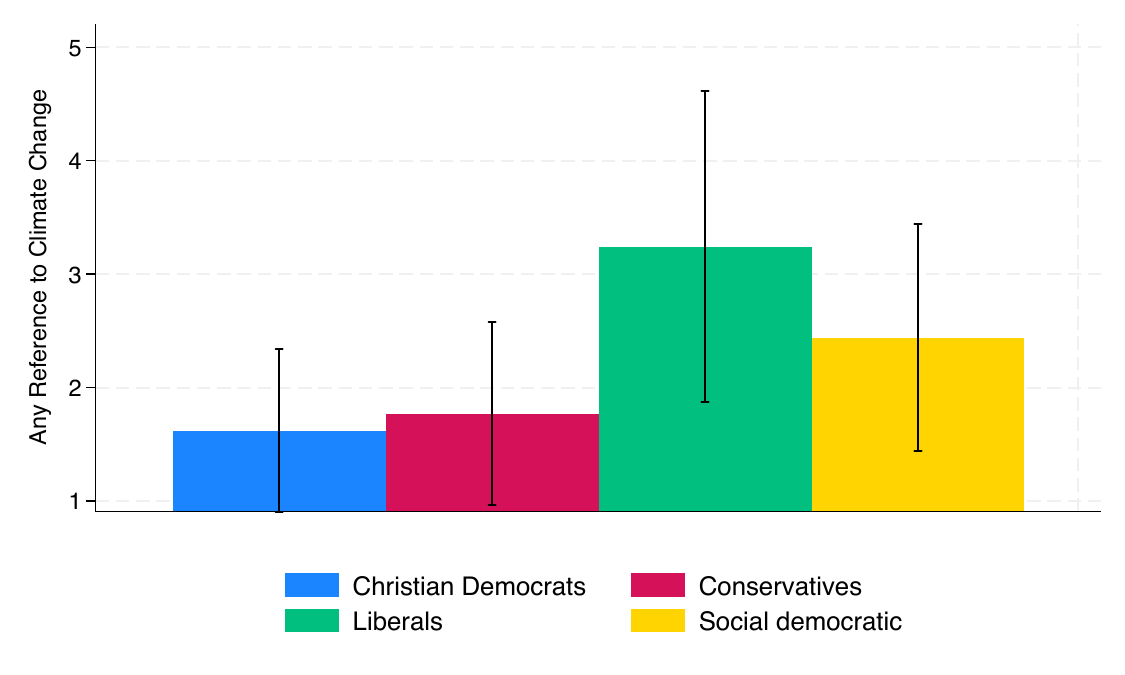}
\label{}
\end{subfigure}\hspace*{\fill}
\begin{subfigure}{0.5\textwidth}
\caption{Positive references}
\includegraphics[width=8cm]{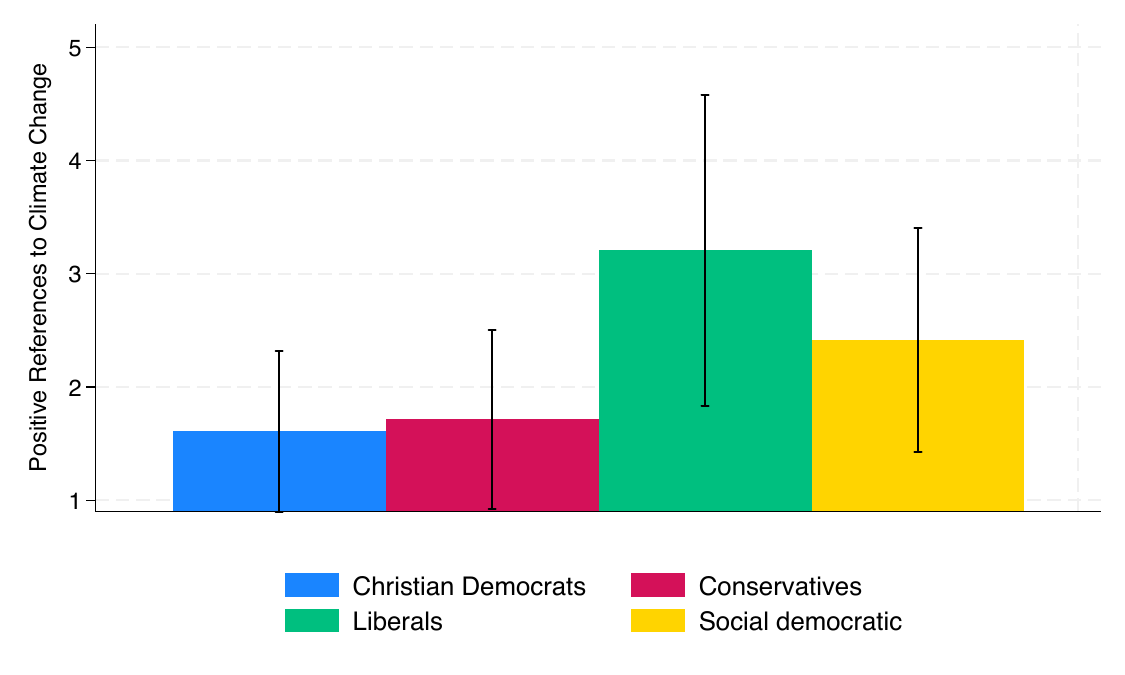}
\label{}
\end{subfigure}
\label{fig_refclimchangbyfamily}
\begin{minipage}{0.9\textwidth} \scriptsize\textbf{Note}: These two figures show the prevalence of references to climate change in the manifestos of mainstream European political parties competing for seats in national parliaments, organised by four party families. The data are from \cite{Schworer2024}. Figure (a) focuses on any reference to climate change, while Figure (b) focuses on positive references (i.e., demands for climate protection, acknowledgement of human-induced climate change, or descriptions of its negative consequences). Vertical lines represent 95\% confidence intervals.
\end{minipage}
\end{center}
\end{figure}

Figure \ref{fig_refclimchangbyfamily} presents the results of the first exercise, showing the proportion of climate change references in party manifestos by family: (a) any reference, and (b) positive references only---i.e., demands for climate protection, recognition of human-made climate change, or descriptions of its negative consequences. Two clear findings emerge: liberals consistently include the largest share of climate change references, followed by social democrats.\footnote{These results align with those of \cite{Schworer2024}.} This evidence is consistent with our hypothesis that these parties integrated climate change into their manifestos and that this integration helped explain their electoral gains. 

Tables \ref{tab_refclimchanglibsocdem} and \ref{tab_refclimchangchrcons} report results from the second exercise, which studies the effect of temperature shocks on the presence of climate change references in manifestos. Specifically, we estimate variations of Equation \eqref{baselinebins_eq}, including party family and country-by-electoral period fixed effects, and report the net effects of temperature shocks for different party families covered in \cite{Schworer2024}. The dependent variables measure climate change references in three ways: (i) any reference, (ii) positive references in favour of climate protection, and (iii) net references (i.e., the difference between pro- and anti-climate protection references). 

Table \ref{tab_refclimchanglibsocdem} focuses on liberal (columns 1-3) and social democratic parties (columns 4-6). It shows positive and statistically significant effects for several temperature bins warmer than the reference interval (10$^{\circ}$C to 25$^{\circ}$C and above 30$^{\circ}$C) across all three dependent variables. For social democrats, results are somewhat weaker: they are statistically significant and larger for the 25$^{\circ}$C–30$^{\circ}$C bin, but insignificant for the 20$^{\circ}$C–25$^{\circ}$C bin. Overall, these findings are consistent with our hypothesis: hotter-than-reference shocks increase the proportion of climate protection references in the manifestos of liberals and social democrats, which are precisely the parties that also gained vote share as a result of such shocks. 

Table \ref{tab_refclimchangchrcons} reports results for Christian democratic (columns 1-3) and conservative parties (columns 4-6). For moderately warm temperatures (10$^{\circ}$C to 25$^{\circ}$C), estimates are small and insignificant. This is consistent with our hypothesis that these parties did not adopt the same strategies as liberals and social democrats, helping to explain their lack of electoral gains. At extremely high temperatures (above 30$^{\circ}$C), however, we find positive effects of similar magnitude to those for liberals and social democrats. This pattern may explain the absence of electoral gains for social democrats and of electoral losses for right-wing parties at these high temperatures:  in this range, liberal gains come primarily at the expense of smaller, less ideologically extreme parties rather than larger, more extreme ones such as right-wing parties, since all major parties place equal emphasis on climate change in their manifestos.

In summary, the evidence presented in this section points to a specific channel through which temperature shocks affect electoral outcomes. Hotter-than-expected shocks above 10$^{\circ}$C lead centrist and electorally vulnerable parties—especially liberals and, to a lesser extent, social democrats—to place greater emphasis on climate protection in their manifestos. These actions, in turn, strengthen their electoral performance in subsequent elections. The broader consequence is a reduction in ideological polarisation and an increase in vote concentration, suggesting a unifying effect of a common external threat.


\subsection{Alternative explanations and limitations}
\label{section_alternative}
We now briefly discuss alternative explanations that could plausibly account for our empirical findings. Specifically, we highlight several indirect channels that are well supported in the related literature \citetext{see, for instance, \citealp{DellJonesOlken2014} and \citealp{CarletonHsiang2026}}. Our aim is not to test these mechanisms empirically, but rather to acknowledge them as limitations of our preferred interpretation and to point to directions for future research.

Temperature shocks may indirectly affect electoral outcomes through several pathways. Economics is one of the most relevant ones: climate variability can influence supply and demand in key markets such as energy and food, reducing agricultural yields, disrupting labour supply and productivity, or affecting trade flows and migration decisions. These effects can spill over into broader macroeconomic outcomes, including GDP growth, poverty, and inequality, and may shift voter preferences depending on how households experience them. 

Beyond the economic impacts, temperature shocks can affect health outcomes, both directly---for instance, by increasing mortality during heatwaves---and indirectly---for instance, by shaping early-life conditions with lasting consequences for well-being. Health shocks of this kind can have strong political implications, as voters may demand more protection from parties that promise stronger social or health-related policies \citep{DeschenesGreenstone2011}. 

Moreover, temperature shocks can influence human behaviour more immediately, for instance by increasing irritability and aggressiveness \citep{Anderson2001, HsiangBurkeMiguel2013}, which in turn can raise levels of interpersonal or intergroup violence. Rising crime and public safety concerns can easily spill over into electoral behaviour, as voters react to heightened insecurity. Taken together, these economic, health, and behavioural channels provide plausible indirect pathways through which temperature shocks may shape electoral outcomes.

Two points are worth noting regarding these mechanisms. On the one hand, they could complement our baseline explanation insofar as changes in economic, health, or behavioural outcomes increase the salience of climate change, thereby favouring parties that emphasise this issue. In this sense, our interpretation would correspond to the final stage of a broader causal process, in which temperature shocks influence intermediate outcomes that ultimately reinforce the political relevance of climate concerns. On the other hand, electoral shifts during periods of economic crises, rising mortality, or heightened crime may instead reflect support for parties associated with greater competence in managing these challenges, independent of their positions on climate change. For example, in contexts of economic downturns induced by temperature shocks, voters may prefer parties offering stronger economic platforms. In the face of excess mortality, parties prioritising social protection may benefit, and in situations of rising crime, parties emphasising security may gain support. If electoral shifts are primarily driven by these responses to temperature-induced crises, rather than by climate concerns per se, then our inability to disentangle these channels represents an important limitation of the study. Identifying the distinct effects of these indirect mechanisms and assessing their empirical relevance remains an important avenue for future research.

\section{Conclusion}
\label{section_conclusion}

This paper studies the impact of short-term temperature shocks on parliamentary electoral outcomes in European regions. Using high-resolution climate data, we find that for the electoral periods from 1989 to 2019, temperature changes could influence voter behaviour in parliamentary elections. In particular, we documented that anomalously warm days reduce ideological polarisation and increase vote concentration. These aggregate patterns reflect a process of electoral convergence towards less radicalised political positions, where centrist parties---particularly liberals and, to a lesser extent, social democrats---gain vote share at the expense of more extreme parties. Our results show that far-right parties lose more votes compared to those at the opposite extreme. Taken together, these results highlight that climate shocks may restructure party systems by pushing voters toward the political centre and reshaping the configuration of electoral competition.

Beyond these demand-side shifts, we also show that some parties strategically adapt their supply of political programs in response to temperature shocks. Liberal and social democratic parties increase the salience of climate change in their manifestos after anomalously warm periods, which likely reinforces their electoral advantage. This suggests that climate change influences both voter behaviour and party strategies, amplifying the political effects of environmental shocks. The broader implication is that climate change may foster a new dimension of electoral competition in Europe, altering the balance between centrists and extremes.

Our findings also come with limitations that point to avenues for further research. On the one hand, our data are limited to European Parliament elections, which, in terms of political ideology, may differ from national or local elections, as well as in voter behaviour. It is also relevant to note that the focus on temperature shocks leaves aside other climatic phenomena, such as floods, droughts, or storms, which may stimulate complementary political responses. Regarding the characterisation of parties, data on party manifestos considers mainly mainstream parties, excluding smaller or emerging political groups that may hold platforms more closely aligned with climate change issues. Finally, it is important to extend the analysis to other electoral arenas and explore the interaction of climate shocks with other pressing electoral and policy issues, such as technological change, social inequality, labour markets, migration, and the dynamics of social media.


\section*{Acknowledgements}
MD and AM acknowledge the financial support of the ST4TE: Strategies for just and equitable transitions in Europe project, funded by the Horizon Europe: Research and Innovation Programme under grant agreement No. 101132559.


\clearpage
\newpage
\bibliographystyle{chicago}
\bibliography{tempelectoral}

\clearpage
\newpage
 \begin{table}[H]
\setlength{\tabcolsep}{15pt}
\begin{center}
\caption {Descriptive Statistics}  \label{tab_descriptivestatistics}
\vspace{-0.3cm}
 \begin{tabular}{l >{\centering\arraybackslash}m{2cm} >{\centering\arraybackslash}m{2cm} >{\centering\arraybackslash}m{2cm}}
\toprule
    & \multicolumn{1}{c}{Obs.} & \multicolumn{1}{c}{Mean}  & \multicolumn{1}{c}{St. dev.}  \\
    \cmidrule(lr){2-2}\cmidrule(lr){3-3} \cmidrule(lr){4-4} 
            &         (1)&         (2)&         (3)\\
\midrule
\emph{\textbf{Temperature Variables}}&            &            &            \\
\# of days with T < -25&        1507&       0.000&       0.000\\
\# of  days with  -25 $\leq$ T < -20&        1507&       0.000&       0.000\\
\# of  days with -20 $\leq$ T < -15&        1507&       0.000&       0.000\\
\# of  days with -15 $\leq$ T < -10&        1507&       0.000&       0.002\\
\# of  days with -10 $\leq$ T < -5&        1507&       0.002&       0.016\\
\# of  days with -5 $\leq$ T < 0&        1507&       0.031&       0.130\\
\# of  days with 0 $\leq$ T < 5&        1507&       0.090&       0.095\\
\# of  days with 5 $\leq$ T < 10&        1507&       0.171&       0.095\\
\# of  days with 10 $\leq$ T < 15&        1507&       0.182&       0.073\\
\# of  days with 15 $\leq$ T < 20&        1507&       0.209&       0.076\\
\# of  days with 20 $\leq$ T < 25&        1507&       0.172&       0.072\\
\# of  days with 25 $\leq$ T < 30&        1507&       0.104&       0.094\\
\# of  days with 30 $\leq$ T < 35&        1507&       0.038&       0.084\\
\# of  days with 35 $\leq$ T&        1507&       0.001&       0.004\\
\emph{\textbf{Party System Outcomes}}&            &            &            \\
Party vote concentration (HHI)&        1508&       0.190&       0.111\\
Party System Ideological polarization&        1514&       0.405&       0.091\\
\emph{\textbf{Vote share by Party Family}}&            &            &            \\
Communist/Socialist&        1514&       0.068&       0.105\\
Green/Ecologist&        1514&       0.083&       0.126\\
Social democratic&        1514&       0.266&       0.161\\
Agrarian    &        1514&       0.014&       0.056\\
Liberal     &        1514&       0.102&       0.148\\
Christian democracy&        1514&       0.141&       0.182\\
Conservative&        1514&       0.203&       0.231\\
Right-wing  &        1514&       0.114&       0.145\\
\emph{\textbf{Demographics}}&            &            &            \\
Population (millions)&        1514&       1.794&       1.532\\
\emph{\textbf{Extreme Weather Events}}&            &            &            \\
Ocurrence of an extreme weather event&        1095&       0.070&       0.256\\
Intensity of the extreme weather events&        1095&       0.359&       6.947\\

\bottomrule
 \multicolumn{4}{p{16cm}}{\scriptsize{\textbf{Note:} This table reports the descriptive statistics of the main variables used in the study. The sample consists of data from 274 regions in Europe at the NUTS-2 level. The temperature variables capture the number of days falling into each specified temperature (T) bin during the twelve months preceding the European Parliament elections. Temperature data are obtained from the Berkeley Earth Foundation dataset, while electoral outcomes come from the ParlGov project. Population data are taken from the Annual Regional Database of the European Commission (ARDECO). Information on extreme weather events is drawn from the International Disaster Database (EM-DAT) and includes two measures: (i) the number of occurrences of any type of extreme weather event in the twelve months prior to each election (“occurrence”), and (ii) the total number of victims resulting from these events, expressed per 100,000 inhabitants (“intensity”).} }
\end{tabular}
\end{center}
\end{table}


\newpage
\begin{table}[H]
\begin{center}
{
\renewcommand{\arraystretch}{1.2}
\caption {Effect on party vote concentration and party system ideological polarisation}  \label{tab_turnHHIpol}
\vspace{-0.3cm}
\small
\centering  \begin{tabular}{l >{\centering\arraybackslash}m{2cm} >{\centering\arraybackslash}m{2cm} >{\centering\arraybackslash}m{2cm} >{\centering\arraybackslash}m{2cm}}
\toprule
& \multicolumn{2}{c}{Party vote} & \multicolumn{2}{c}{Ideological}\\
    & \multicolumn{2}{c}{Concentration} & \multicolumn{2}{c}{polarisation}\\
  \cmidrule[0.2pt](l){2-3}\cmidrule[0.2pt](l){4-5}
& (1)& (2)& (3) & (4)    \\   
\bottomrule
-15 $\leq$ T < -10  &       1.093\sym{***}&       1.140\sym{***}&      -0.110         &      -0.092         \\
                    &     (0.380)         &     (0.391)         &     (0.292)         &     (0.294)         \\
-10 $\leq$ T < -5   &       0.659         &       0.674         &      -0.063         &      -0.070         \\
                    &     (0.502)         &     (0.507)         &     (0.201)         &     (0.200)         \\
-5 $\leq$ T < 0     &       0.355\sym{**} &       0.360\sym{**} &       0.036         &       0.035         \\
                    &     (0.158)         &     (0.158)         &     (0.119)         &     (0.116)         \\
0 $\leq$ T < 5      &       0.101\sym{*}  &       0.110\sym{**} &       0.087         &       0.082         \\
                    &     (0.054)         &     (0.053)         &     (0.080)         &     (0.078)         \\
5 $\leq$ T < 10     &       0.000         &       0.000         &       0.000         &       0.000         \\
                    &         (.)         &         (.)         &         (.)         &         (.)         \\
10 $\leq$ T < 15    &       0.064         &       0.060         &      -0.227\sym{***}&      -0.221\sym{***}\\
                    &     (0.053)         &     (0.054)         &     (0.083)         &     (0.078)         \\
15 $\leq$ T < 20    &       0.219\sym{***}&       0.203\sym{***}&      -0.424\sym{***}&      -0.420\sym{***}\\
                    &     (0.067)         &     (0.071)         &     (0.122)         &     (0.117)         \\
20 $\leq$ T < 25    &       0.254\sym{***}&       0.237\sym{***}&      -0.502\sym{***}&      -0.499\sym{***}\\
                    &     (0.070)         &     (0.075)         &     (0.133)         &     (0.128)         \\
25 $\leq$ T < 30    &       0.366\sym{***}&       0.343\sym{***}&      -0.638\sym{***}&      -0.633\sym{***}\\
                    &     (0.078)         &     (0.081)         &     (0.159)         &     (0.155)         \\
30 $\leq$ T < 35    &       0.432\sym{***}&       0.410\sym{***}&      -0.619\sym{***}&      -0.618\sym{***}\\
                    &     (0.100)         &     (0.091)         &     (0.180)         &     (0.179)         \\
35 $\leq$ T         &       0.422         &       0.428\sym{*}  &      -0.812\sym{***}&      -0.681\sym{**} \\
                    &     (0.258)         &     (0.253)         &     (0.313)         &     (0.268)         \\
R-sq                &       0.969         &       0.969         &       0.931         &       0.932         \\
Observations        &        1483         &        1483         &        1483         &        1483         \\

\hline
Controls & No & Yes & No & Yes \\
\bottomrule
\multicolumn{5}{p{11.5cm}}{\scriptsize{\textbf{Note:} All columns report the estimates from Equation~\eqref{baselinebins_eq}, based on data from 274 regions in Europe at the NUTS-2 level during the twelve months preceding the European Parliament elections. The dependent variable in columns (1) and (2) is party vote concentration, measured by the Herfindahl–Hirschman Index (see Equation~\eqref{eq:hhi}). The dependent variable in columns (3) and (4) is party system ideological polarisation (see Equation~\eqref{eq:polarisation}). All models include region fixed effects and country–year fixed effects. Columns (2) and (4) additionally control for precipitation, extreme weather events, and the natural logarithm of population. Robust standard errors (in parentheses) are clustered at the region level. * denotes statistical significance at the 10\% level, ** at the 5\% level, and *** at the 1\% level.} }
\end{tabular}
}
\end{center}
\end{table}


\newpage

\begin{table}[H]
\begin{center}
{
\renewcommand{\arraystretch}{1.2}
\setlength{\tabcolsep}{5pt}
\caption {Effect on vote share by party family}  
\label{tab_idelologypartyfamily}
\vspace{-0.3cm}
\small
\centering  \begin{tabular}{lccccccccc}
\toprule
  &  \multicolumn{1}{c}{com} & \multicolumn{1}{c}{eco} & \multicolumn{1}{c}{soc} & \multicolumn{1}{c}{agr} & \multicolumn{1}{c}{lib} & \multicolumn{1}{c}{chr} & \multicolumn{1}{c}{con}& \multicolumn{1}{c}{right} \\
  \cmidrule[0.2pt](l){2-2}\cmidrule[0.2pt](l){3-3}\cmidrule[0.2pt](l){4-4}\cmidrule[0.2pt](l){5-5}\cmidrule[0.2pt](l){6-6}\cmidrule[0.2pt](l){7-7}\cmidrule[0.2pt](l){8-8}\cmidrule[0.2pt](l){9-9}
& (1)& (2)& (3)& (4)& (5)& (6)& (7)& (8) \\ 
\midrule
-15$\leq$T<-10           &       0.649\sym{***}&      -0.357         &      -0.176         &       0.651\sym{**} &      -0.425         &      -0.327         &       0.063         &      -0.199         \\
                    &     (0.228)         &     (0.292)         &     (0.460)         &     (0.286)         &     (0.339)         &     (0.331)         &     (0.459)         &     (0.352)         \\
-10$\leq$T<-5            &       0.615\sym{*}  &      -0.048         &      -0.311         &      -0.530\sym{***}&      -0.026         &      -0.021         &      -0.002         &       0.234         \\
                    &     (0.319)         &     (0.160)         &     (0.361)         &     (0.151)         &     (0.296)         &     (0.163)         &     (0.223)         &     (0.205)         \\
-5$\leq$T<0              &       0.081         &       0.046         &      -0.143         &      -0.119\sym{**} &      -0.140         &       0.040         &       0.015         &       0.088         \\
                    &     (0.104)         &     (0.074)         &     (0.158)         &     (0.054)         &     (0.128)         &     (0.130)         &     (0.198)         &     (0.147)         \\
0$\leq$T<5               &      -0.066         &       0.091\sym{*}  &      -0.163\sym{*}  &      -0.077\sym{***}&      -0.125\sym{**} &      -0.009         &       0.055         &       0.258\sym{**} \\
                    &     (0.077)         &     (0.051)         &     (0.099)         &     (0.026)         &     (0.062)         &     (0.059)         &     (0.122)         &     (0.129)         \\
5$\leq$T<10              &       0.000         &       0.000         &       0.000         &       0.000         &       0.000         &       0.000         &       0.000         &       0.000         \\
                    &         (.)         &         (.)         &         (.)         &         (.)         &         (.)         &         (.)         &         (.)         &         (.)         \\
10$\leq$T<15             &      -0.089         &       0.096\sym{**} &       0.130         &      -0.039         &       0.178\sym{**} &       0.001         &       0.033         &      -0.254\sym{**} \\
                    &     (0.082)         &     (0.048)         &     (0.088)         &     (0.027)         &     (0.072)         &     (0.047)         &     (0.106)         &     (0.102)         \\
15$\leq$T<20             &      -0.189\sym{*}  &       0.115         &       0.302\sym{***}&      -0.130\sym{**} &       0.266\sym{***}&       0.027         &       0.025         &      -0.358\sym{***}\\
                    &     (0.098)         &     (0.075)         &     (0.097)         &     (0.051)         &     (0.080)         &     (0.046)         &     (0.148)         &     (0.117)         \\
20$\leq$T<25             &      -0.252\sym{*}  &       0.072         &       0.395\sym{***}&      -0.154\sym{**} &       0.266\sym{**} &       0.075         &      -0.077         &      -0.318\sym{**} \\
                    &     (0.134)         &     (0.069)         &     (0.130)         &     (0.062)         &     (0.111)         &     (0.055)         &     (0.136)         &     (0.134)         \\
25$\leq$T<30             &      -0.258         &      -0.096         &       0.603\sym{***}&      -0.200\sym{**} &       0.383\sym{***}&       0.063         &      -0.156         &      -0.348\sym{**} \\
                    &     (0.175)         &     (0.062)         &     (0.215)         &     (0.078)         &     (0.129)         &     (0.054)         &     (0.151)         &     (0.167)         \\
30$\leq$T<35             &      -0.172         &      -0.342\sym{***}&       0.607\sym{**} &      -0.303\sym{***}&       0.558\sym{***}&       0.099         &       0.009         &      -0.459\sym{**} \\
                    &     (0.208)         &     (0.111)         &     (0.280)         &     (0.110)         &     (0.183)         &     (0.072)         &     (0.178)         &     (0.226)         \\
35$\leq$T                &       0.264         &      -1.317         &       0.451         &      -0.436\sym{**} &       0.819\sym{***}&      -0.022         &       0.258         &      -0.011         \\
                    &     (0.419)         &     (0.854)         &     (0.909)         &     (0.190)         &     (0.290)         &     (0.147)         &     (0.635)         &     (0.316)         \\
R-sq                &       0.916         &       0.969         &       0.935         &       0.948         &       0.958         &       0.984         &       0.970         &       0.936         \\
Observations        &        1483         &        1483         &        1483         &        1483         &        1483         &        1483         &        1483         &        1483         \\

\bottomrule
\multicolumn{9}{p{15cm}}{\scriptsize{\textbf{Note:} 
 All columns report estimates from Equation \eqref{baselinebins_eq}, using data from 274 European regions at the NUTS-2 level during the twelve months preceding the European Parliament elections. The dependent variable in each column is the vote share of the corresponding party family in the European Parliament elections. Party family abbreviations are: \emph{com} = Communist/Socialist, \emph{eco} = Green/Ecologist, \emph{soc} = Social Democratic, \emph{agr} = Agrarian, \emph{lib} = Liberal, \emph{chr} = Christian Democratic, \emph{con} = Conservative, and \emph{right} = Right-Wing. All models include region fixed effects, country–year fixed effects, and controls for precipitation, population, and both the presence and intensity of extreme weather events. Robust standard errors (in parentheses) are clustered at the regional level. * denotes statistical significance at the 10\% level, ** at the 5\% level, and *** at the 1\% level.} }
\end{tabular}
}
\end{center}
\end{table}

\newpage


\begin{table}[H]
\begin{center}
{
\renewcommand{\arraystretch}{1.2}
\caption {Effect on party vote concentration and party system ideological polarisation by historic average temperature}  \label{Table_histavgtemp_HHIPOL}
\vspace{-0.3cm}
\small
\centering  \begin{tabular}{lcccc}
\toprule
  &  \multicolumn{2}{c}{Party vote concentration} & \multicolumn{2}{c}{Ideological polarisation}\\
  \cmidrule[0.2pt](l){2-3}\cmidrule[0.2pt](l){4-5}
&  \multicolumn{1}{c}{Above-median} & \multicolumn{1}{c}{Below-median}  &  \multicolumn{1}{c}{Above-median} & \multicolumn{1}{c}{Below-median}   \\
&  \multicolumn{1}{c}{historic temp} & \multicolumn{1}{c}{historic temp}  &  \multicolumn{1}{c}{historic temp} & \multicolumn{1}{c}{historic temp}   \\
\cmidrule[0.2pt](l){2-2}\cmidrule[0.2pt](l){3-3}\cmidrule[0.2pt](l){4-4}\cmidrule[0.2pt](l){5-5}
& (1)& (2)& (3) & (4)   \\   
\midrule
-15 $\leq$ T < -10  &                     &       1.090\sym{***}&                     &      -0.188         \\
                    &                     &     (0.377)         &                     &     (0.298)         \\
-10 $\leq$ T < -5   &                     &       0.018         &                     &      -0.004         \\
                    &                     &     (0.207)         &                     &     (0.178)         \\
-5 $\leq$ T < 0     &       0.039         &       0.200         &       0.166         &      -0.013         \\
                    &     (0.178)         &     (0.163)         &     (0.468)         &     (0.127)         \\
0 $\leq$ T < 5      &       0.059         &       0.056         &       0.100         &       0.059         \\
                    &     (0.081)         &     (0.065)         &     (0.179)         &     (0.089)         \\
5 $\leq$ T < 10     &       0.000         &       0.000         &       0.000         &       0.000         \\
                    &         (.)         &         (.)         &         (.)         &         (.)         \\
10 $\leq$ T < 15    &       0.084         &       0.065         &      -0.258\sym{**} &      -0.197\sym{*}  \\
                    &     (0.070)         &     (0.075)         &     (0.114)         &     (0.118)         \\
15 $\leq$ T < 20    &       0.207\sym{*}  &       0.204\sym{**} &      -0.453\sym{**} &      -0.371\sym{**} \\
                    &     (0.107)         &     (0.087)         &     (0.186)         &     (0.173)         \\
20 $\leq$ T < 25    &       0.204\sym{*}  &       0.300\sym{***}&      -0.569\sym{***}&      -0.426\sym{**} \\
                    &     (0.111)         &     (0.098)         &     (0.205)         &     (0.203)         \\
25 $\leq$ T < 30    &       0.250\sym{**} &       0.422\sym{***}&      -0.715\sym{***}&      -0.571\sym{**} \\
                    &     (0.124)         &     (0.113)         &     (0.226)         &     (0.258)         \\
30 $\leq$ T < 35    &       0.295\sym{***}&       0.120         &      -0.721\sym{***}&      -0.152         \\
                    &     (0.112)         &     (0.204)         &     (0.248)         &     (0.232)         \\
35 $\leq$ T         &       0.240         &                     &      -0.868\sym{***}&                     \\
                    &     (0.231)         &                     &     (0.278)         &                     \\
R-sq                &       0.964         &       0.981         &       0.934         &       0.934         \\
Observations        &         733         &         718         &         733         &         718         \\

\bottomrule
\multicolumn{5}{p{13.5cm}}{\scriptsize{\textbf{Note:} All columns report estimates from Equation \eqref{baselinebins_eq}, using data from 274 European regions at the NUTS-2 level during the twelve months preceding the European Parliament elections. In columns (1) and (2), the dependent variable is party vote concentration, measured by the Herfindahl–Hirschman Index (see Equation~\eqref{eq:hhi}). In columns (3) and (4), the dependent variable is party system ideological polarisation (see Equation~\eqref{eq:polarisation}). Columns (1) and (3) are restricted to historically warm regions (above the median temperature), while columns (2) and (4) are restricted to historically cold regions (below the median temperature). All models include region fixed effects, country–year fixed effects, and controls for precipitation, population, and both the number and intensity of extreme weather events. Robust standard errors (in parentheses) are clustered at the regional level. * denotes statistical significance at the 10\% level, ** at the 5\% level, and *** at the 1\% level. } }
\end{tabular}
}
\end{center}
\end{table}


\newpage


\begin{table}[H]
\begin{center}
{
\renewcommand{\arraystretch}{1.1}
\setlength{\tabcolsep}{5pt}
\caption {Effect on vote share by party family: regions with above-median historic temperature}  \label{tab_idelologypartyfamily_high}
\vspace{-0.3cm}
\small
\centering  \begin{tabular}{lccccccccc}
\toprule
  &  \multicolumn{1}{c}{com} & \multicolumn{1}{c}{eco} & \multicolumn{1}{c}{soc} & \multicolumn{1}{c}{agr} & \multicolumn{1}{c}{lib} & \multicolumn{1}{c}{chr} & \multicolumn{1}{c}{con}& \multicolumn{1}{c}{right} \\
  \cmidrule[0.2pt](l){2-2}\cmidrule[0.2pt](l){3-3}\cmidrule[0.2pt](l){4-4}\cmidrule[0.2pt](l){5-5}\cmidrule[0.2pt](l){6-6}\cmidrule[0.2pt](l){7-7}\cmidrule[0.2pt](l){8-8}\cmidrule[0.2pt](l){9-9}
& (1)& (2)& (3)& (4)& (5)& (6)& (7)& (8) \\    
\midrule
-10$\leq$T<-5            &       0.643         &      -0.688         &      26.832\sym{***}&      -2.878\sym{**} &      -4.209         &      -0.361         &       0.316         &     -20.162\sym{***}\\
                    &     (1.999)         &     (2.746)         &     (4.697)         &     (1.369)         &     (4.215)         &     (1.091)         &     (3.553)         &     (3.272)         \\
-5$\leq$T<0              &       0.355\sym{*}  &       0.417\sym{*}  &      -0.463         &       0.127         &      -0.320         &       0.171\sym{*}  &      -0.109         &      -0.181         \\
                    &     (0.194)         &     (0.241)         &     (0.335)         &     (0.283)         &     (0.358)         &     (0.094)         &     (0.325)         &     (0.170)         \\
0$\leq$T<5               &      -0.020         &      -0.107         &      -0.393\sym{**} &      -0.111\sym{*}  &      -0.055         &       0.012         &       0.282         &       0.401         \\
                    &     (0.161)         &     (0.081)         &     (0.170)         &     (0.065)         &     (0.107)         &     (0.080)         &     (0.206)         &     (0.278)         \\
5$\leq$T<10              &       0.000         &       0.000         &       0.000         &       0.000         &       0.000         &       0.000         &       0.000         &       0.000         \\
                    &         (.)         &         (.)         &         (.)         &         (.)         &         (.)         &         (.)         &         (.)         &         (.)         \\
10$\leq$T<15             &      -0.014         &       0.054         &       0.120         &      -0.036         &       0.161\sym{*}  &       0.042         &      -0.001         &      -0.263\sym{*}  \\
                    &     (0.112)         &     (0.051)         &     (0.115)         &     (0.046)         &     (0.092)         &     (0.059)         &     (0.131)         &     (0.152)         \\
15$\leq$T<20             &      -0.081         &       0.175         &       0.195         &      -0.116         &       0.330\sym{***}&       0.028         &      -0.198         &      -0.378\sym{**} \\
                    &     (0.119)         &     (0.123)         &     (0.129)         &     (0.086)         &     (0.112)         &     (0.060)         &     (0.202)         &     (0.168)         \\
20$\leq$T<25             &      -0.354\sym{*}  &       0.108         &       0.410\sym{**} &      -0.159         &       0.423\sym{**} &      -0.005         &      -0.160         &      -0.253         \\
                    &     (0.206)         &     (0.130)         &     (0.179)         &     (0.105)         &     (0.176)         &     (0.068)         &     (0.172)         &     (0.179)         \\
25$\leq$T<30             &      -0.332         &      -0.218\sym{***}&       0.618\sym{**} &      -0.193\sym{*}  &       0.446\sym{**} &       0.025         &      -0.004         &      -0.405\sym{**} \\
                    &     (0.248)         &     (0.081)         &     (0.305)         &     (0.105)         &     (0.172)         &     (0.060)         &     (0.175)         &     (0.166)         \\
30$\leq$T<35             &      -0.221         &      -0.460\sym{***}&       0.733\sym{**} &      -0.298\sym{**} &       0.499\sym{***}&       0.070         &       0.146         &      -0.553\sym{**} \\
                    &     (0.258)         &     (0.113)         &     (0.365)         &     (0.150)         &     (0.191)         &     (0.073)         &     (0.217)         &     (0.222)         \\
35$\leq$T                &       0.053         &      -1.844\sym{*}  &       0.915         &      -0.432\sym{**} &       0.586\sym{*}  &      -0.053         &       0.992         &      -0.153         \\
                    &     (0.393)         &     (0.944)         &     (1.055)         &     (0.195)         &     (0.314)         &     (0.142)         &     (0.664)         &     (0.329)         \\
R-sq                &       0.929         &       0.980         &       0.951         &       0.901         &       0.956         &       0.976         &       0.973         &       0.922         \\
Observations        &         733         &         733         &         733         &         733         &         733         &         733         &         733         &         733         \\

\bottomrule
\multicolumn{9}{p{15cm}}{\scriptsize{\textbf{Note:} 
All columns report estimates from Equation \eqref{baselinebins_eq}, using data from 274 European regions at the NUTS-2 level during the twelve months preceding the European Parliament elections. All columns are restricted to historically warm regions (above the median temperature). The dependent variable in each column is the vote share of the corresponding party family in the European Parliament elections. Party family abbreviations are: \emph{com} = Communist/Socialist, \emph{eco} = Green/Ecologist, \emph{soc} = Social Democratic, \emph{agr} = Agrarian, \emph{lib} = Liberal, \emph{chr} = Christian Democratic, \emph{con} = Conservative, and \emph{right} = Right-Wing. All models include region fixed effects, country–year fixed effects, and controls for precipitation, population, and both the presence and intensity of extreme weather events. Robust standard errors (in parentheses) are clustered at the regional level. * denotes statistical significance at the 10\% level, ** at the 5\% level, and *** at the 1\% level.} }
\end{tabular}
}
\end{center}
\end{table}

\newpage


\begin{table}[H]
\begin{center}
{
\renewcommand{\arraystretch}{1.2}
\setlength{\tabcolsep}{5pt}
\caption {Effect on vote share by party family: regions with below-median historic temperature}  \label{tab_idelologypartyfamily_low}
\vspace{-0.3cm}
\small
\centering  \begin{tabular}{lccccccccc}
\bottomrule
  &  \multicolumn{1}{c}{com} & \multicolumn{1}{c}{eco} & \multicolumn{1}{c}{soc} & \multicolumn{1}{c}{agr} & \multicolumn{1}{c}{lib} & \multicolumn{1}{c}{chr} & \multicolumn{1}{c}{con}& \multicolumn{1}{c}{right} \\\cmidrule[0.2pt](l){2-2}\cmidrule[0.2pt](l){3-3}\cmidrule[0.2pt](l){4-4}\cmidrule[0.2pt](l){5-5}\cmidrule[0.2pt](l){6-6}\cmidrule[0.2pt](l){7-7}\cmidrule[0.2pt](l){8-8}\cmidrule[0.2pt](l){9-9}
& (1)& (2)& (3)& (4)& (5)& (6)& (7)& (8) \\   
\midrule
-15$\leq$T<-10           &       0.178         &      -0.297         &       0.402         &       0.819\sym{***}&      -0.404         &       0.102         &      -0.276         &      -0.426         \\
                    &     (0.352)         &     (0.305)         &     (0.632)         &     (0.260)         &     (0.423)         &     (0.432)         &     (0.488)         &     (0.479)         \\
-10$\leq$T<-5            &       0.451         &       0.156         &       0.015         &      -0.325\sym{***}&      -0.323         &       0.231         &      -0.232         &       0.078         \\
                    &     (0.348)         &     (0.145)         &     (0.419)         &     (0.109)         &     (0.232)         &     (0.224)         &     (0.252)         &     (0.263)         \\
-5$\leq$T<0              &      -0.056         &       0.241\sym{***}&       0.117         &      -0.025         &      -0.224         &       0.254         &      -0.291         &      -0.013         \\
                    &     (0.168)         &     (0.087)         &     (0.240)         &     (0.016)         &     (0.173)         &     (0.176)         &     (0.195)         &     (0.202)         \\
0$\leq$T<5               &      -0.140         &       0.199\sym{***}&       0.055         &      -0.007         &      -0.221\sym{**} &       0.079         &      -0.115         &       0.154         \\
                    &     (0.109)         &     (0.058)         &     (0.132)         &     (0.011)         &     (0.093)         &     (0.066)         &     (0.141)         &     (0.137)         \\
5$\leq$T<10              &       0.000         &       0.000         &       0.000         &       0.000         &       0.000         &       0.000         &       0.000         &       0.000         \\
                    &         (.)         &         (.)         &         (.)         &         (.)         &         (.)         &         (.)         &         (.)         &         (.)         \\
10$\leq$T<15             &      -0.138         &       0.149\sym{**} &       0.119         &      -0.026\sym{*}  &       0.077         &       0.013         &       0.104         &      -0.272\sym{*}  \\
                    &     (0.097)         &     (0.067)         &     (0.166)         &     (0.015)         &     (0.089)         &     (0.075)         &     (0.163)         &     (0.155)         \\
15$\leq$T<20             &      -0.224         &       0.175\sym{**} &       0.295         &      -0.066\sym{**} &       0.124         &       0.185\sym{**} &       0.018         &      -0.389\sym{*}  \\
                    &     (0.139)         &     (0.084)         &     (0.215)         &     (0.027)         &     (0.098)         &     (0.086)         &     (0.191)         &     (0.220)         \\
20$\leq$T<25             &      -0.134         &       0.232\sym{**} &       0.306         &      -0.071\sym{*}  &       0.097         &       0.333\sym{***}&      -0.308         &      -0.342         \\
                    &     (0.205)         &     (0.100)         &     (0.269)         &     (0.036)         &     (0.126)         &     (0.108)         &     (0.205)         &     (0.284)         \\
25$\leq$T<30             &      -0.129         &       0.204\sym{*}  &       0.503         &      -0.082\sym{*}  &       0.191         &       0.308\sym{**} &      -0.656\sym{***}&      -0.201         \\
                    &     (0.265)         &     (0.118)         &     (0.337)         &     (0.049)         &     (0.156)         &     (0.120)         &     (0.245)         &     (0.374)         \\
30$\leq$T<35             &       0.586\sym{*}  &      -0.390         &      -0.409         &      -0.113\sym{*}  &       0.675\sym{*}  &       0.064         &      -0.313         &      -0.027         \\
                    &     (0.330)         &     (0.280)         &     (0.541)         &     (0.058)         &     (0.383)         &     (0.276)         &     (0.312)         &     (0.453)         \\
R-sq                &       0.880         &       0.927         &       0.924         &       0.997         &       0.967         &       0.989         &       0.966         &       0.957         \\
Observations        &         718         &         718         &         718         &         718         &         718         &         718         &         718         &         718         \\

\bottomrule
\multicolumn{9}{p{15cm}}{\scriptsize{\textbf{Note:} 
All columns report estimates from Equation \eqref{baselinebins_eq}, using data from 274 European regions at the NUTS-2 level during the twelve months preceding the European Parliament elections. All columns are restricted to historically cold regions (below the median temperature). The dependent variable in each column is the vote share of the corresponding party family in the European Parliament elections. Party family abbreviations are: \emph{com} = Communist/Socialist, \emph{eco} = Green/Ecologist, \emph{soc} = Social Democratic, \emph{agr} = Agrarian, \emph{lib} = Liberal, \emph{chr} = Christian Democratic, \emph{con} = Conservative, and \emph{right} = Right-Wing. All models include region fixed effects, country–year fixed effects, and controls for precipitation, population, and both the presence and intensity of extreme weather events. Robust standard errors (in parentheses) are clustered at the regional level. * denotes statistical significance at the 10\% level, ** at the 5\% level, and *** at the 1\% level.} }
\end{tabular}
}
\end{center}
\end{table}


\begin{table}[H]
\begin{center}
{
\renewcommand{\arraystretch}{1.2}
\setlength{\tabcolsep}{8pt}
\caption {Effect on parties’ vote share by Liberal and Social Democratic strength in previous election}  \label{tab_idelologypartyfamily_strenghtpast}
\vspace{-0.3cm}
\small
\centering  \begin{tabular}{lcccccc}
\toprule
  &  \multicolumn{3}{c}{Above-median liberal \& soc. dem.} & \multicolumn{3}{c}{Below-median liberal \& soc. dem.}\\
 &  \multicolumn{3}{c}{vote share in previous election} & \multicolumn{3}{c}{vote share in previous election}\\ 
 \cmidrule[0.2pt](l){2-4}\cmidrule[0.2pt](l){5-7} 
&  \multicolumn{1}{c}{soc} & \multicolumn{1}{c}{lib}  &  \multicolumn{1}{c}{right} & \multicolumn{1}{c}{soc} & \multicolumn{1}{c}{lib}  &  \multicolumn{1}{c}{right}    \\\cmidrule[0.2pt](l){2-2}\cmidrule[0.2pt](l){3-3}\cmidrule[0.2pt](l){4-4}\cmidrule[0.2pt](l){5-5}\cmidrule[0.2pt](l){6-6}\cmidrule[0.2pt](l){7-7}
& (1)& (2)& (3) & (4) & (5) & (6)   \\   
\midrule
-15$\leq$T<-10           &       4.164\sym{**} &       7.558\sym{***}&       1.712         &       0.689         &      -0.488         &      -0.658         \\
                    &     (1.622)         &     (1.754)         &     (2.978)         &     (0.690)         &     (0.643)         &     (0.405)         \\
-10$\leq$T<-5            &       0.148         &      -2.300\sym{***}&       0.327         &       0.253         &      -0.364         &      -0.332         \\
                    &     (0.595)         &     (0.633)         &     (0.782)         &     (0.449)         &     (0.371)         &     (0.242)         \\
-5$\leq$T<0              &      -0.141         &      -0.404\sym{*}  &       0.354         &       0.022         &      -0.425         &      -0.137         \\
                    &     (0.269)         &     (0.219)         &     (0.444)         &     (0.298)         &     (0.312)         &     (0.193)         \\
0$\leq$T<5               &       0.130         &      -0.221\sym{**} &       0.320         &      -0.368\sym{**} &      -0.231\sym{*}  &       0.097         \\
                    &     (0.144)         &     (0.105)         &     (0.334)         &     (0.182)         &     (0.131)         &     (0.094)         \\
5$\leq$T<10              &       0.000         &       0.000         &       0.000         &       0.000         &       0.000         &       0.000         \\
                    &         (.)         &         (.)         &         (.)         &         (.)         &         (.)         &         (.)         \\
10$\leq$T<15             &       0.093         &      -0.056         &      -0.160         &       0.041         &       0.389\sym{**} &      -0.290\sym{***}\\
                    &     (0.143)         &     (0.090)         &     (0.246)         &     (0.179)         &     (0.157)         &     (0.103)         \\
15$\leq$T<20             &       0.178         &       0.039         &       0.006         &       0.183         &       0.629\sym{***}&      -0.421\sym{***}\\
                    &     (0.183)         &     (0.100)         &     (0.202)         &     (0.197)         &     (0.225)         &     (0.155)         \\
20$\leq$T<25             &       0.020         &       0.033         &       0.120         &       0.428\sym{*}  &       0.792\sym{***}&      -0.607\sym{***}\\
                    &     (0.183)         &     (0.129)         &     (0.233)         &     (0.230)         &     (0.303)         &     (0.212)         \\
25$\leq$T<30             &       0.018         &       0.289         &       0.018         &       0.713\sym{***}&       0.886\sym{**} &      -0.729\sym{***}\\
                    &     (0.233)         &     (0.181)         &     (0.225)         &     (0.272)         &     (0.364)         &     (0.244)         \\
30$\leq$T<35             &      -0.192         &       0.859\sym{**} &       0.252         &       0.975\sym{***}&       1.240\sym{***}&      -0.930\sym{***}\\
                    &     (0.296)         &     (0.342)         &     (0.304)         &     (0.337)         &     (0.474)         &     (0.328)         \\
35$\leq$T                &      -2.427\sym{***}&       1.469\sym{**} &       0.714         &       4.076\sym{***}&       0.899         &      -0.875         \\
                    &     (0.726)         &     (0.645)         &     (0.523)         &     (1.289)         &     (1.043)         &     (0.653)         \\
R-sq                &       0.972         &       0.965         &       0.929         &       0.938         &       0.950         &       0.985         \\
Observations        &         570         &         570         &         570         &         571         &         571         &         571         \\

\bottomrule
\multicolumn{7}{p{13.8cm}}{\scriptsize{\textbf{Note:} All columns report estimates from Equation \eqref{baselinebins_eq}, using data from 274 European regions at the NUTS-2 level during the twelve months preceding the European Parliament elections. The dependent variable in each column is the vote share of the corresponding party family, where \emph{soc} = Social Democratic, \emph{lib} = Liberal, and \emph{right} = Right-Wing. Columns (1) to (3) are restricted to regions where liberal and social democratic parties performed well in the previous elections (vote share above the median), while columns (4) to (6) are restricted to regions where these parties performed poorly (vote share below the median). All models include region fixed effects, country–year fixed effects, and controls for precipitation, population, and both the presence and intensity of extreme weather events. Robust standard errors (in parentheses) are clustered at the regional level. * denotes statistical significance at the 10\% level, ** at the 5\% level, and *** at the 1\% level.} }
\end{tabular}
}
\end{center}
\end{table}



\begin{table}[H]
\begin{center}
{
\renewcommand{\arraystretch}{1.2}
\setlength{\tabcolsep}{9pt}
\caption {Effect on the inclusion of climate change issues in Liberal and Social Democratic parties' manifestos}  \label{tab_refclimchanglibsocdem}
\vspace{-0.3cm}
\small
\centering  \begin{tabular}{lcccccc}
\toprule
  &  \multicolumn{6}{c}{Dependent variable: references to climate change in manifestos of}\\\cmidrule[0.2pt](l){2-7}
 &  \multicolumn{3}{c}{Liberal parties}  &  \multicolumn{3}{c}{Social Democratic parties} \\\cmidrule[0.2pt](l){2-4}\cmidrule[0.2pt](l){5-7}
&  \multicolumn{1}{c}{Any}  &   \multicolumn{1}{c}{Positive} & \multicolumn{1}{c}{Net} &  \multicolumn{1}{c}{Any}  &   \multicolumn{1}{c}{Positive} & \multicolumn{1}{c}{Net} \\
 &  \multicolumn{1}{c}{reference} &   \multicolumn{1}{c}{references} & \multicolumn{1}{c}{references}   &  \multicolumn{1}{c}{reference} &   \multicolumn{1}{c}{references} & \multicolumn{1}{c}{references}   \\
 \cmidrule[0.2pt](l){2-2}\cmidrule[0.2pt](l){3-3}\cmidrule[0.2pt](l){4-4}\cmidrule[0.2pt](l){5-5}\cmidrule[0.2pt](l){6-6}\cmidrule[0.2pt](l){7-7}
  & (1)& (2)& (3) & (4)& (5)& (6) \\
\midrule
(-15$\leq$T<-10)         &      -1.805**         &      -1.747**         &      -1.752**         &       6.899**         &       6.902**         &       6.923**         \\
                    &     (0.686)         &     (0.672)         &     (0.670)         &     (2.441)         &     (2.429)         &     (2.452)         \\
(-10$\leq$T<-5)          &       0.032         &      -0.008         &      -0.012         &      -1.481***         &      -1.480***         &      -1.485***         \\
                    &     (0.239)         &     (0.238)         &     (0.237)         &     (0.259)         &     (0.250)         &     (0.254)         \\
(-5$\leq$T<0)            &       0.592*         &       0.577*         &       0.576*         &       0.456***         &       0.462***         &       0.471***         \\
                    &     (0.277)         &     (0.270)         &     (0.267)         &     (0.090)         &     (0.087)         &     (0.089)         \\
(0$\leq$T<5)             &      -0.185**         &      -0.184**         &      -0.183**         &      -0.164         &      -0.160         &      -0.162         \\
                    &     (0.064)         &     (0.064)         &     (0.064)         &     (0.093)         &     (0.091)         &     (0.091)         \\
(5$\leq$T<10)            &       0.000         &       0.000         &       0.000         &       0.000         &       0.000         &       0.000         \\
                    &     (.)         &     (.)         &     (.)         &     (.)         &     (.)         &     (.)         \\
(10$\leq$T<15)           &       0.523*         &       0.508*         &       0.506*         &       0.225         &       0.225         &       0.230         \\
                    &     (0.257)         &     (0.249)         &     (0.246)         &     (0.131)         &     (0.127)         &     (0.126)         \\
(15$\leq$T<20)           &       0.279**         &       0.272**         &       0.270**         &       0.223**         &       0.220**         &       0.221**         \\
                    &     (0.084)         &     (0.082)         &     (0.081)         &     (0.068)         &     (0.065)         &     (0.065)         \\
(20$\leq$T<25)           &       0.698**         &       0.665**         &       0.657**         &       0.098         &       0.092         &       0.092         \\
                    &     (0.262)         &     (0.254)         &     (0.250)         &     (0.105)         &     (0.101)         &     (0.100)         \\
(25$\leq$T<30)           &       0.198         &       0.163         &       0.156         &       0.204***         &       0.196***         &       0.197***         \\
                    &     (0.130)         &     (0.126)         &     (0.125)         &     (0.042)         &     (0.042)         &     (0.041)         \\
(30$\leq$T<35)           &       0.998*         &       1.078*         &       1.105*         &       0.083         &       0.082         &       0.080         \\
                    &     (0.492)         &     (0.475)         &     (0.468)         &     (0.069)         &     (0.068)         &     (0.068)         \\
(35$\leq$T)              &      10.950**         &      11.002**         &      10.897**         &      22.484***         &      22.451***         &      22.426***         \\
                    &     (3.835)         &     (3.861)         &     (3.855)         &     (4.177)         &     (3.999)         &     (4.089)         \\
R-sq                &       0.673         &       0.665         &       0.661         &       0.816         &       0.812         &       0.812         \\
Observations        &          99         &          99         &          99         &          71         &          71         &          71         \\

\bottomrule
\multicolumn{7}{p{15cm}}{\scriptsize{\textbf{Note:} All columns report estimates from Equation \eqref{baselinebins_eq}, but using party-level data by year, covering 27 mainstream political parties in eight European countries during national parliamentary elections held between 1995 and 2022. The dependent variable in all columns measures the presence of references to climate change in party manifestos, based on data from \cite{Schworer2024}. In columns (1) and (4), the dependent variable is the presence of any reference to climate change. In columns (2) and (5), it is the presence of positive references (i.e., demands for climate protection, acknowledgement of human-induced climate change, or descriptions of its negative consequences). In columns (3) and (6), it is the presence of net references (i.e., the difference between pro- and anti-climate protection references). Reported coefficients capture the net effect of temperature shocks for liberal parties (columns (1)–(3)) and social democratic parties (columns (4)–(6)). All models include party family fixed effects, country–year fixed effects, and controls for precipitation. Robust standard errors (in parentheses) are clustered at the regional level.  * denotes statistical significance at the 10\% level, ** at the 5\% level, and *** at the 1\% level.} }
\end{tabular}
}
\end{center}
\end{table}

\begin{table}[H]
\begin{center}
{
\renewcommand{\arraystretch}{1.2}
\setlength{\tabcolsep}{9pt}
\caption {Effect on the inclusion of climate change issues in Christian Democratic and Conservative Parties' manifestos}  \label{tab_refclimchangchrcons}
\vspace{-0.3cm}
\small
\centering  \begin{tabular}{lcccccc}
\toprule
  &  \multicolumn{6}{c}{Dependent variable: references to climate change in manifestos of}\\\cmidrule[0.2pt](l){2-7}
 &  \multicolumn{3}{c}{Christian Democratic parties}  &  \multicolumn{3}{c}{Conservative parties} \\\cmidrule[0.2pt](l){2-4}\cmidrule[0.2pt](l){5-7}
&  \multicolumn{1}{c}{Any}  &   \multicolumn{1}{c}{Positive} & \multicolumn{1}{c}{Net} &  \multicolumn{1}{c}{Any}  &   \multicolumn{1}{c}{Positive} & \multicolumn{1}{c}{Net} \\
 &  \multicolumn{1}{c}{reference} &   \multicolumn{1}{c}{references} & \multicolumn{1}{c}{references}   &  \multicolumn{1}{c}{reference} &   \multicolumn{1}{c}{references} & \multicolumn{1}{c}{references}   \\
 \cmidrule[0.2pt](l){2-2}\cmidrule[0.2pt](l){3-3}\cmidrule[0.2pt](l){4-4}\cmidrule[0.2pt](l){5-5}\cmidrule[0.2pt](l){6-6}\cmidrule[0.2pt](l){7-7}
  & (1)& (2)& (3) & (4)& (5)& (6) \\   
  \midrule
(-15$\leq$T<-10)         &       3.392         &       3.467         &       3.455         &       1.383         &       1.329         &       1.328         \\
                    &     (1.976)         &     (1.955)         &     (1.959)         &     (1.223)         &     (1.232)         &     (1.232)         \\
(-10$\leq$T<-5)          &      -0.189         &      -0.205         &      -0.200         &       0.261         &       0.296         &       0.297         \\
                    &     (0.169)         &     (0.168)         &     (0.172)         &     (0.192)         &     (0.193)         &     (0.193)         \\
(-5$\leq$T<0)            &      -0.239         &      -0.230         &      -0.221         &      -0.340*         &      -0.345*         &      -0.338*         \\
                    &     (0.184)         &     (0.182)         &     (0.184)         &     (0.166)         &     (0.164)         &     (0.164)         \\
(0$\leq$T<5)             &      -0.032         &      -0.033         &      -0.035         &       0.088         &       0.100         &       0.099         \\
                    &     (0.066)         &     (0.066)         &     (0.066)         &     (0.063)         &     (0.063)         &     (0.063)         \\
(5$\leq$T<10)            &       0.000         &       0.000         &       0.000         &       0.000         &       0.000         &       0.000         \\
                    &     (.)         &     (.)         &     (.)         &     (.)         &     (.)         &     (.)         \\
(10$\leq$T<15)           &      -0.046         &      -0.047         &      -0.039         &      -0.082         &      -0.097         &      -0.092         \\
                    &     (0.290)         &     (0.289)         &     (0.289)         &     (0.142)         &     (0.142)         &     (0.142)         \\
(15$\leq$T<20)           &       0.184         &       0.180         &       0.183         &       0.120         &       0.101         &       0.102         \\
                    &     (0.174)         &     (0.174)         &     (0.174)         &     (0.129)         &     (0.130)         &     (0.129)         \\
(20$\leq$T<25)           &       0.162         &       0.152         &       0.157         &      -0.135         &      -0.172         &      -0.171         \\
                    &     (0.254)         &     (0.254)         &     (0.253)         &     (0.121)         &     (0.121)         &     (0.121)         \\
(25$\leq$T<30)           &       0.112         &       0.107         &       0.110         &       0.138**         &       0.127**         &       0.128**         \\
                    &     (0.171)         &     (0.170)         &     (0.168)         &     (0.041)         &     (0.042)         &     (0.042)         \\
(30$\leq$T<35)           &      -0.643         &      -0.667         &      -0.669         &      -0.246***         &      -0.260***         &      -0.261***         \\
                    &     (0.409)         &     (0.405)         &     (0.401)         &     (0.030)         &     (0.029)         &     (0.029)         \\
(35$\leq$T)              &      13.767***         &      13.873***         &      13.794***         &      17.266***         &      17.496***         &      17.403***         \\
                    &     (2.110)         &     (2.095)         &     (2.119)         &     (0.999)         &     (0.986)         &     (0.989)         \\
R-sq                &       0.648         &       0.640         &       0.638         &       0.660         &       0.655         &       0.653         \\
Observations        &          99         &          99         &          99         &          99         &          99         &          99         \\

\bottomrule
\multicolumn{7}{p{15cm}}{\scriptsize{\textbf{Note:} All columns report estimates from Equation \eqref{baselinebins_eq}, but using party-level data by year, covering 27 mainstream political parties in eight European countries during national parliamentary elections held between 1995 and 2022. The dependent variable in all columns measures the presence of references to climate change in party manifestos, based on data from \cite{Schworer2024}. In columns (1) and (4), the dependent variable is the presence of any reference to climate change. In columns (2) and (5), it is the presence of positive references (i.e., demands for climate protection, acknowledgement of human-induced climate change, or descriptions of its negative consequences). In columns (3) and (6), it is the presence of net references (i.e., the difference between pro- and anti-climate protection references). Reported coefficients capture the net effect of temperature shocks for Christian democratic parties (columns (1)–(3)) and conservative parties (columns (4)–(6)). All models include party family fixed effects, country–year fixed effects, and controls for precipitation. Robust standard errors (in parentheses) are clustered at the regional level.  * denotes statistical significance at the 10\% level, ** at the 5\% level, and *** at the 1\% level.} }
\end{tabular}
}
\end{center}
\end{table}


\clearpage
\appendix
\section{Appendix: Additional figures and tables}
\label{webapp1}

\setcounter{table}{0}
\setcounter{figure}{0}
\setcounter{subsection}{0}
\renewcommand{\thefigure}{\Alph{section}\arabic{figure}}
\renewcommand{\thetable}{\Alph{section}\arabic{table}}


\begin{figure}[H]
\begin{center}
\caption{Spatial and temporal heterogeneity of temperatures in European regions}
   \begin{subfigure}{0.5\textwidth}
\includegraphics[width=7cm]{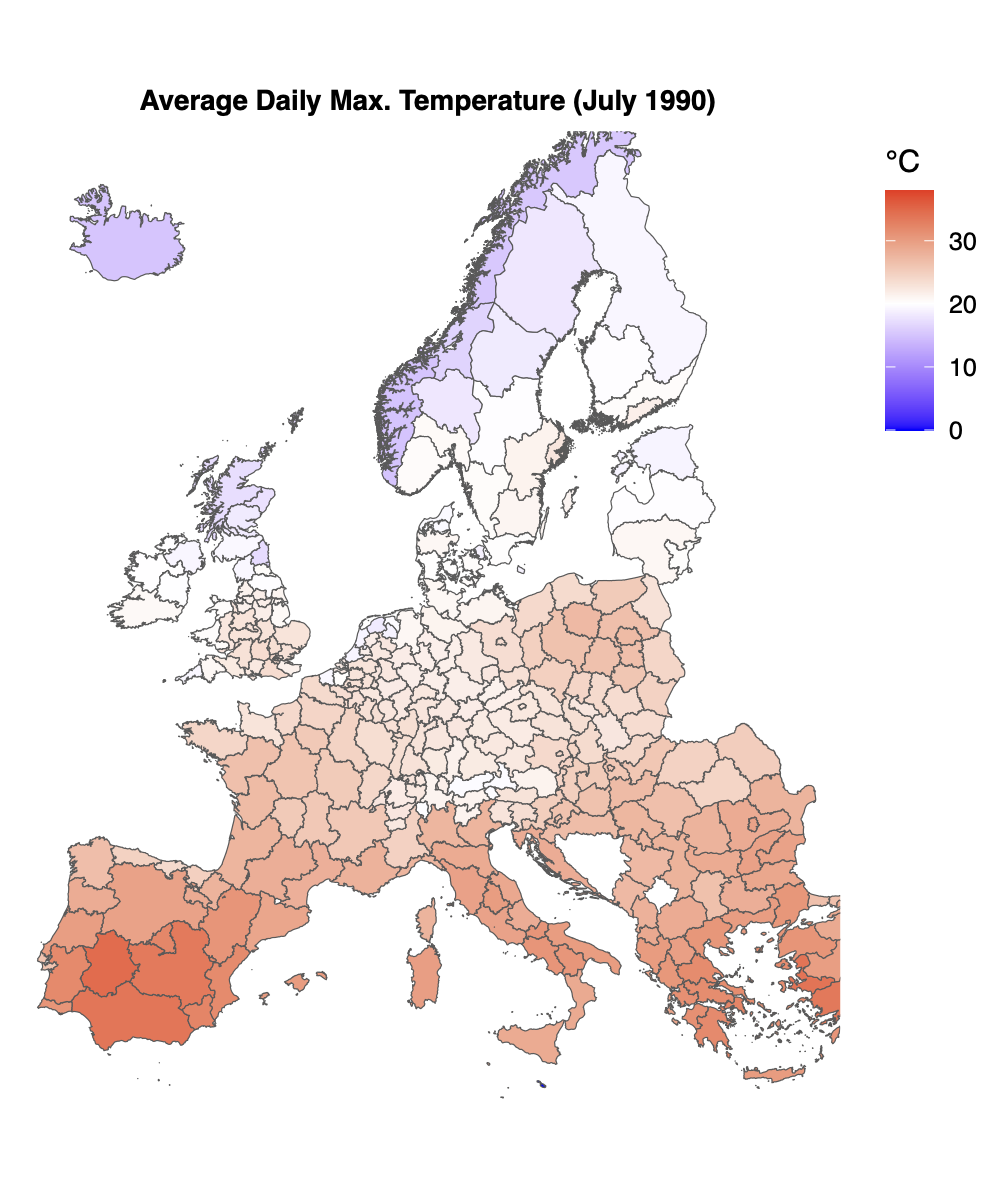}
 \caption{July 1990}
 \label{fig:first}
\end{subfigure}\hspace*{\fill}
   \begin{subfigure}{0.5\textwidth}
\includegraphics[width=7cm]{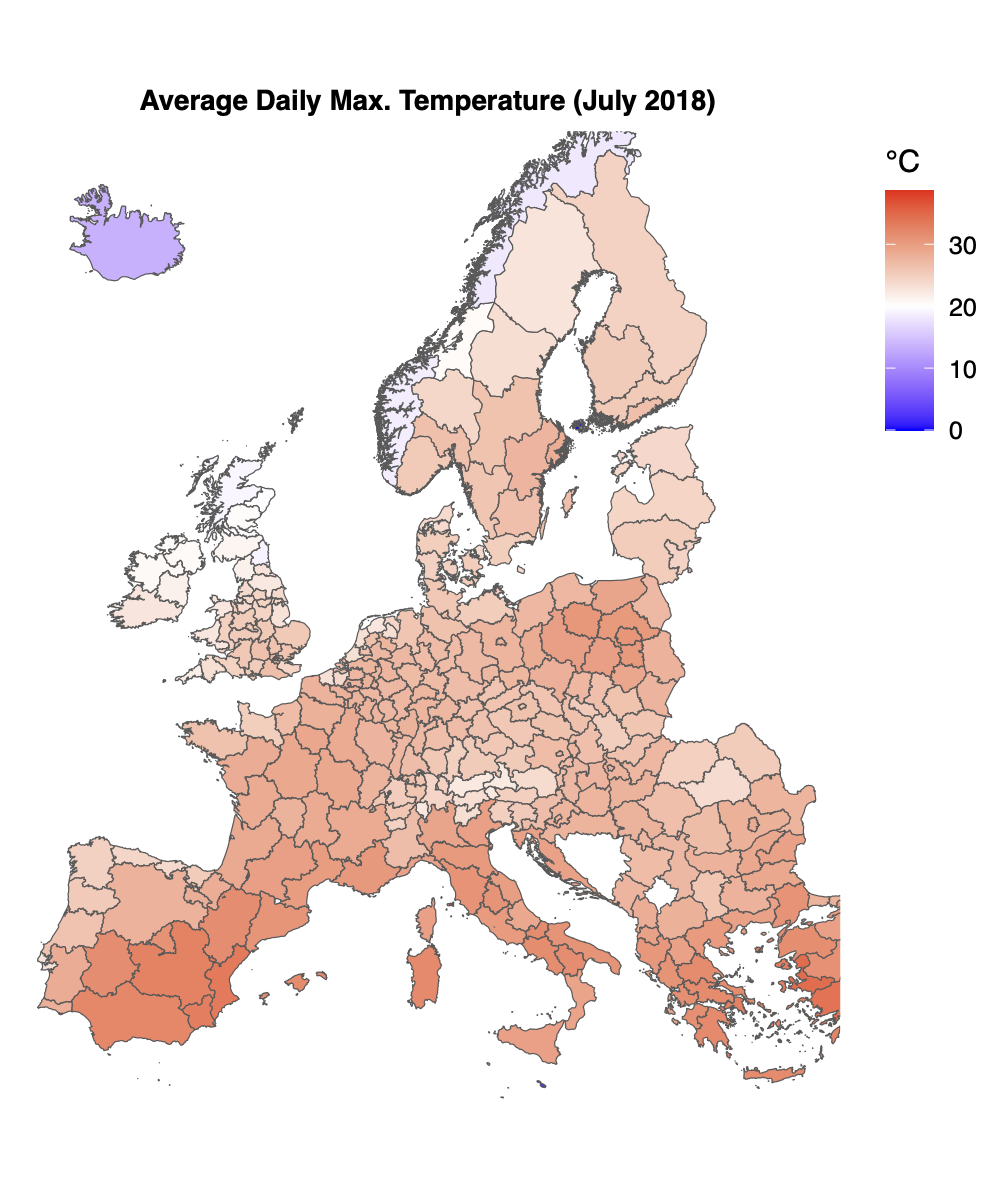}
 \caption{July 2018}
 \label{fig:first}
\end{subfigure}
   \begin{subfigure}{0.5\textwidth}
\includegraphics[width=7cm]{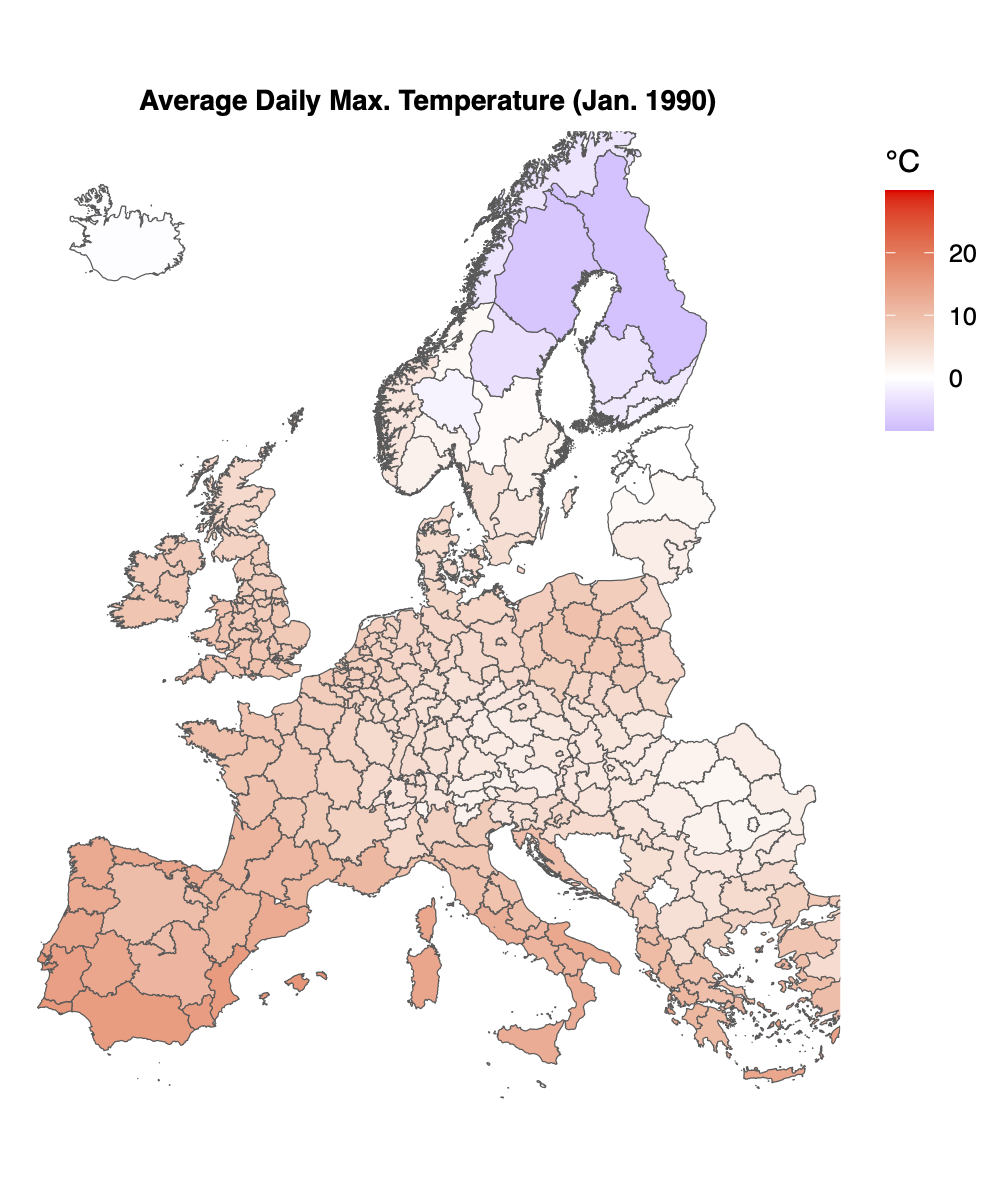}
 \caption{Jan 1990}
 \label{fig:first}
\end{subfigure}\hspace*{\fill}
   \begin{subfigure}{0.5\textwidth}
\includegraphics[width=7cm]{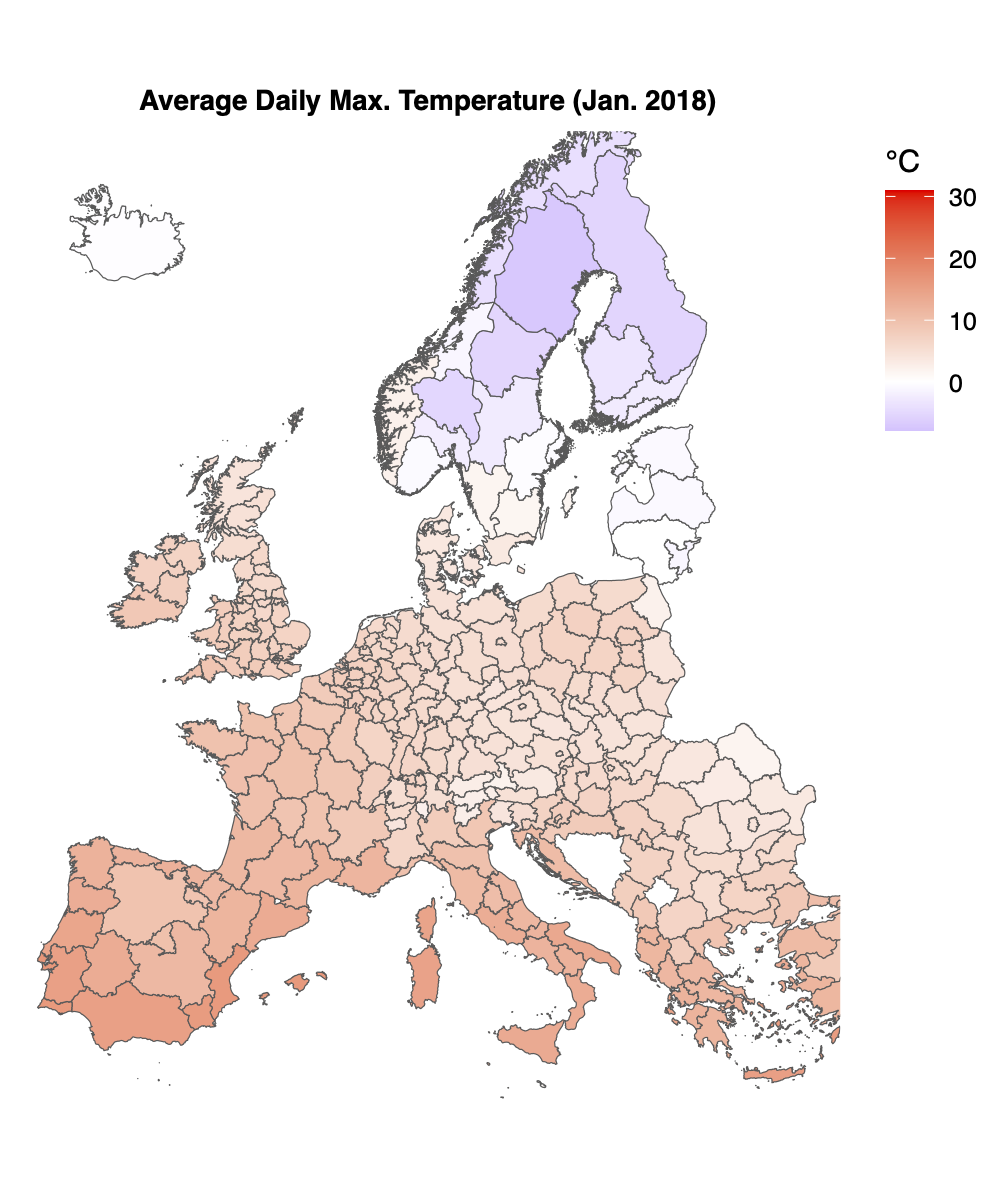}
 \caption{Jan 2018}
 \label{fig:first}
\end{subfigure}
\label{fig_map}
\begin{minipage}{0.9\textwidth} \scriptsize\textbf{Note}: These figures illustrate the spatial heterogeneity of temperatures across European regions for different years in our sample. Each figure reports the average daily maximum temperature. Figure (a) shows July 1990, Figure (b) July 2018, Figure (c) January 1990, and Figure (d) January 2018.
\end{minipage}
\end{center}
\end{figure}




\begin{table}[H]
\begin{center}
{
\renewcommand{\arraystretch}{1}
\setlength{\tabcolsep}{4pt}
\caption {Effect on vote concentration, ideological polarisation and vote share: standard panel analysis}  
\vspace{-0.3cm}
\label{table_panelalloutcomes}
\resizebox{0.9\textwidth}{!}{\renewcommand{\arraystretch}{1.1}
\small
\centering  \begin{tabular}{lcccccccccc}
\toprule
  & \multicolumn{1}{c}{HHI} & \multicolumn{1}{c}{Pol.}  & \multicolumn{1}{c}{com} & \multicolumn{1}{c}{eco} & \multicolumn{1}{c}{soc} & \multicolumn{1}{c}{agr} & \multicolumn{1}{c}{lib} & \multicolumn{1}{c}{chr} & \multicolumn{1}{c}{con}& \multicolumn{1}{c}{right} \\
  \cmidrule[0.2pt](l){2-2}\cmidrule[0.2pt](l){3-3}\cmidrule[0.2pt](l){4-4}\cmidrule[0.2pt](l){5-5}\cmidrule[0.2pt](l){6-6}\cmidrule[0.2pt](l){7-7}\cmidrule[0.2pt](l){8-8}\cmidrule[0.2pt](l){9-9}\cmidrule[0.2pt](l){10-10}\cmidrule[0.2pt](l){11-11}
& (1)& (2)& (3)& (4)& (5)& (6)& (7)& (8)& (9) & (10)  \\   
\midrule
Sum of all temp. coeff.&       0.003         &      -0.013**         &       0.002         &      -0.005         &       0.015         &       0.002         &       0.012**          &       0.003         &      -0.005         &      -0.021**         \\
                    &     (0.005)         &     (0.005)         &     (0.009)         &     (0.004)         &     (0.010)         &     (0.003)         &     (0.006)        &     (0.004)         &     (0.010)         &     (0.008)         \\
R-sq                &       0.971         &       0.939         &       0.923         &       0.971         &       0.937         &       0.950         &       0.961         &       0.985         &       0.971         &       0.937         \\
Observations        &        1471         &        1471         &        1471         &        1471         &        1471         &        1471         &        1471         &        1471         &        1471         &        1471         \\

\bottomrule
\multicolumn{11}{p{17.5cm}}{\scriptsize{\textbf{Note:} All columns report estimates from the specification described in footnote \ref{footnote_panelspec}. All models include precipitation controls, region fixed effects, year fixed effects, and country-year fixed effects. The dependent variable in column (1) is party vote concentration, in column (2) ideological polarisation, and in columns (3) through (10) the vote share of the following party families: Christian Democratic (\emph{chr}), Green/Ecologist (\emph{eco}), Liberal (\emph{lib}), Right-Wing (\emph{right}), Social Democratic (\emph{soc}), Conservative (\emph{con}), Communist/Socialist (\emph{com}), and Agrarian (\emph{agr}), respectively. Robust standard errors (in parentheses) are clustered by region. * denotes statistical significance at the 10\% level, ** at the 5\% level, and *** at the 1\% level.} }
\end{tabular}
}}
\end{center}
\end{table}

\end{document}